\documentclass{aa}
\usepackage{amsmath,graphicx,amssymb}
\usepackage[varg]{txfonts}
\usepackage{natbib}
\usepackage{pst-eps}
\usepackage{subfloat}
\usepackage{stfloats}
\bibpunct{(}{)}{,}{a}{}{,}

\newcommand{\zav}[1]{\left(#1\right)}
\newcommand{\hzav}[1]{\left[#1\right]}

\newlength\staretab
\newcommand{\Teff}{\mbox{$T_\mathrm{eff}$}}
\newcommand\de{\text{d}}
\newcommand\x[1]{\ensuremath{#1_\text{X}}}
\newcommand\lx{\ensuremath{\x L}}
\newcommand\vnek{\ensuremath{\varv_\infty}}
\newcommand\msr{\ensuremath{M_\odot\,\text{yr}^{-1}}}
\newcommand\kms{\ensuremath{\text{km}\,\text{s}^{-1}}}

\newcommand\cc{\ensuremath{C_\text{c}}}
\newcommand{\vel}{{\varv}}

\begin{document}

\title{Stellar wind models of central stars of planetary nebulae}

\author{J.~Krti\v{c}ka\inst{1} \and J. Kub\'at\inst{2} \and
I.~Krti\v{c}kov\'a\inst{1}}


\institute{\'Ustav teoretick\'e fyziky a astrofyziky, Masarykova univerzita,
           Kotl\'a\v rsk\' a 2, CZ-611\,37 Brno, Czech
           Republic
           \and
           Astronomick\'y \'ustav, Akademie v\v{e}d \v{C}esk\'e
           republiky, Fri\v{c}ova 298, CZ-251 65 Ond\v{r}ejov, Czech Republic}

\date{Received}

\abstract{Fast line-driven stellar winds play an
important role in the evolution of planetary nebulae, even though they are relatively weak.}{We provide global (unified)
hot star wind models of central stars of planetary nebulae. The models predict
wind structure including the mass-loss rates, terminal velocities, and emergent
fluxes from basic stellar parameters.}{We applied our wind code for parameters
corresponding to evolutionary stages between the asymptotic giant branch and white
dwarf phases for a star with a final mass of $0.569\,M_\odot$. We study the influence
of metallicity and wind inhomogeneities (clumping) on the wind
properties.}{Line-driven winds appear very early after the star leaves the
asymptotic giant branch (at the latest for $T_\text{eff}\approx10\,$kK) and fade
away at the white dwarf cooling track (below $T_\text{eff}=105\,$kK). Their
mass-loss rate mostly scales with the stellar luminosity and, consequently, the
mass-loss rate only varies slightly during the transition from the red to the blue part
of the Hertzsprung-Russell diagram. There are the following two exceptions to the monotonic
behavior: a bistability jump at around $20\,$kK, where the mass-loss rate
decreases by a factor of a few (during evolution) due to a change in iron
ionization, and an additional maximum at about $T_\text{eff}=40-50\,$kK. On the
other hand, the terminal velocity increases from about a few hundreds of \kms\ to
a few thousands of \kms\ during the transition as a result of stellar radius
decrease. The wind terminal velocity also significantly increases at the
bistability jump.  Derived wind parameters reasonably agree with observations.
The effect of clumping is stronger at the hot side of the bistability jump than
at the cool side.}{Derived fits to wind parameters can be used in evolutionary
models and in studies of planetary nebula formation. A predicted bistability jump
in mass-loss rates can cause the appearance of an additional shell of planetary
nebula.}


\titlerunning{Stellar wind models of central stars of planetary nebulae}

\authorrunning{J.~Krti\v{c}ka et al.}
\maketitle

\section{Introduction}

 The transition between the asymptotic giant branch (AGB) and white dwarf (WD)
phases belongs to one of the most prominent phases of stellar evolution.
After leaving the AGB phase, the stars for which the expansion time of the AGB
envelope is comparable to the transition time-scale create planetary nebulae
\citep{kwoksan}. During this phase, the hot central star of planetary nebula (CSPN)
ionizes the material left over from previous evolutionary phases and fast wind of the
central star collides with slow material lost during previous evolution
\citep{pekny2}. This creates a magnificent nebula that is visible from extragalactic
distances \citep[e.g.,][]{herci,rimcc}.

Although evolutionary tracks after leaving the AGB phase are rather simple
\citep{vasiwo,blok}, their detailed properties are sensitive to processes
occurring during previous evolutionary phases \citep{milbe}. The detailed
treatment of opacities, nuclear reaction rates, and AGB mass-loss affects the
predictions of evolutionary models. Yet there is another process that appears
after leaving the AGB branch that may affect the evolutionary timescales and
appearance of CSPNe: a stellar wind in post-AGB and CSPN phases.

Stellar winds of CSPNe are driven in a similar way as winds of their more
massive counterparts, that is, by light absorption and scattering in the lines of heavier
elements, including carbon, nitrogen, oxygen, and iron \citep{btpaupuv}. The
mass-loss rate of hot star winds mainly depends on the stellar luminosity
\citep{cak} and since the winds are primarily driven by heavier elements, also on
metallicity \citep{vikolamet,metuje}. On the other hand, wind terminal velocity,
which is the wind velocity at large distances from the star, mainly depends on
the surface escape speed \citep{cak,lsl}.

Stellar winds of CSPN are accessible for observation at intermediate evolutionary
stages between AGB and WD. Observational probes of CSPN winds include
ultraviolet (UV) P~Cygni line profiles and H$\alpha$ emission
\citep{kuplan,herbian} and possibly also X-rays \citep{chu,guesbor}. The
observational studies enabled us to derive the wind parameters, that is, the wind
mass-loss rate and the terminal velocity.

On the other hand, the stellar winds in the phase immediately following the AGB
phase (the post-AGB phase) and at the top of the WD cooling track are difficult
to probe observationally due to a fast speed of evolution and weak winds,
respectively. However, the winds can be very important even in these elusive
stages. Strong winds in a post-AGB phase may affect the duration of this phase
\citep{milbe} and winds in hot WDs may prevent gravitational settling
\citep{unbu}, thus explaining near-solar abundances in these stars
\citep{wrk17,wrk18}.

To better understand the final evolutionary stages of solar-mass stars, we
provide theoretical models of their stellar winds. We used our own wind code to
predict wind properties of the stars that just left the AGB phase, CSPNe, and
the stars that started their final descend along the WD cooling sequence.

\section{Description of the CMF wind models}

\begin{table*}
\caption{Atoms and ions included in the NLTE and line force calculations. The
column "Level" lists a total number of individual levels and superlevels. The
column "Data" gives the source of data for NLTE ($a$) and for line radiative
force calculations ($b$) in a format of $a/b$. The column "Range" gives the
temperature range (in kK) in which the given ion is included in calculations.
A blank item means that the ion is always included.}
\label{prvky}
\centering
\begin{tabular}{*{4}{l@{\hspace{0.5mm}}r@{\hspace{2.5mm}}c@{\hspace{2.5mm}}c}}
\hline
Ion & Level & Data & Range & Ion & Level & Data & Range & Ion & Level & Data & Range & Ion & Level & Data & Range\\
\hline
\ion{H}{i}   &  9& 1/2 &     & \ion{Ne}{i}   & 15& 2/3 & <25 & \ion{Al}{i}   & 10& 1/3 & <25 & \ion{Ar}{ii}  & 21& 2/3 & <25\\
\ion{H}{ii}  &  1& 1   &     & \ion{Ne}{ii}  & 15& 1/3 & <50 & \ion{Al}{ii}  & 16& 1/3 & <50 & \ion{Ar}{iii} & 25&2/3,15& <80\\
\ion{He}{i}  & 14& 1/3 & <90 & \ion{Ne}{iii} & 14&1/3,15&    & \ion{Al}{iii} & 14& 2/3 & <99 & \ion{Ar}{iv}  & 19& 2/3 & <80\\
\ion{He}{ii} & 14& 1/3 &     & \ion{Ne}{iv}  & 12&1/3,15&    & \ion{Al}{iv}  & 14&2/15& <99 & \ion{Ar}{v}   & 16&11/3 & <80\\
\ion{He}{iii}&  1& 1   &     & \ion{Ne}{v}   & 17&6/3,15&    & \ion{Al}{v}   & 16&2/3,15&<99 & \ion{Ar}{vi}  & 11& 2/3 & <80\\
\ion{C}{i}   & 26& 1/3 & <25 & \ion{Ne}{vi}  & 11&2/3,15&    & \ion{Al}{vi}  & 1&2/4&25--98 & \ion{Ar}{vii} &  1& 2/3 & <80\\
\ion{C}{ii}  & 14&1/3,4& <50 & \ion{Ne}{vii} & 16& 7/4 & >25 & \ion{Si}{ii}  & 12&1/3,15& <50 & \ion{Ca}{ii}  & 16&2/3,15& <50\\
\ion{C}{iii} & 23&1/3,4&     & \ion{Ne}{viii}& 16& 8,12/4&>80& \ion{Si}{iii} & 12& 1/3 &     & \ion{Ca}{iii} & 14& 2/3 & <50\\
\ion{C}{iv}  & 25&1/3,4,15&  & \ion{Ne}{ix}  &  1& 2/4 &>99  & \ion{Si}{iv}  & 13&1/3,15&    & \ion{Ca}{iv}  & 20&2/3,15& <50\\
\ion{C}{v}   & 11&5/4  &     & \ion{Na}{i}   & 14& 2/3 & <25 & \ion{Si}{v}   & 15&2/15&      & \ion{Ca}{v}   & 22&2/3,15& <50\\
\ion{C}{vi}  &  1& 2/2 & >25 & \ion{Na}{ii}  & 13& 2/3 & <50 & \ion{Si}{vi}  & 17& 2/4 & >25 & \ion{Ca}{vi}  &  1& 2/3 & <50\\
\ion{N}{i}   & 21& 1/3 & <25 & \ion{Na}{iii} & 14&2/3,15&<90 & \ion{Si}{vii} & 25& 2/4 & >50 & \ion{Fe}{i}   & 38& 1/3 & <25\\
\ion{N}{ii}  & 14&1/3,4,15&<50&\ion{Na}{iv}  & 18&2/3,15&<90 & \ion{Si}{viii}&  1& 2/4 & >99& \ion{Fe}{ii}  & 35& 1/3 & <25\\ 
\ion{N}{iii} & 32&1/3,4,15&  & \ion{Na}{v}   & 16&2/3,15&<90 & \ion{P }{ii}  & 10& 9/3 & <25 & \ion{Fe}{iii} & 29& 2,16/3 & <50\\
\ion{N}{iv}  & 23&1/3,4,15&  & \ion{Na}{vi}  & 1& 2/3,15&25--90& \ion{P }{iii} & 16&9/3,15& <80 & \ion{Fe}{iv} & 32& 17,18/3 & <90\\
\ion{N}{v}   & 16&1/3,4&     & \ion{Mg}{ii}  & 14& 1/3 & <25 & \ion{P }{iv}  & 17& 9/3 &     & \ion{Fe}{v}   & 30&13/3\\
\ion{N}{vi}  & 15&5/3,4,15&  & \ion{Mg}{iii} & 14&2/15&      & \ion{P }{v}   & 21&9/3,15&     & \ion{Fe}{vi}  & 27& 2,19/3\\
\ion{N}{vii} &  1& 2/4 & >25 & \ion{Mg}{iv}  & 14&2/3,15&    & \ion{P }{vi}  & 14&9/15&     & \ion{Fe}{vii} & 29&14/3\\
\ion{O}{i}   & 12&1/3,4& <25 & \ion{Mg}{v}   & 13&2/3,15&    & \ion{P }{vii} &  1& 9/3 & >25 & \ion{Fe}{viii}& 22& 2/15& >50\\
\ion{O}{ii}  & 50&1/3,15&<50 & \ion{Mg}{vi}  & 12&2/3,15& >25& \ion{S }{ii}  & 14&1/3,15& <50 & \ion{Fe}{ix}  & 34& 2/15& >50\\
\ion{O}{iii} & 29&1/3,4&     & \ion{Mg}{vii} & 12& 6/4 & >50 & \ion{S }{iii} & 10&1/3,15& <50 & \ion{Fe}{x}   & 28& 2/15& >80\\
\ion{O}{iv}  & 39&1/3,4,15&  & \ion{Mg}{viii}&  1& 2/4 & >80 & \ion{S }{iv}  & 18& 2/3 & <50 & \ion{Fe}{xi}  &  1& 2/3 & >90\\  
\ion{O}{v}   & 14&1/3,4,15&  &               &   &     &     & \ion{S }{v}   & 14&10/3 & <50 & \ion{Ni}{ii}  & 36& 1/3 & <25\\  
\ion{O}{vi}  & 20&1/3,4,15&  &               &   &     &     & \ion{S }{vi}  & 16&1/3,15& <50 & \ion{Ni}{iii} & 36& 1/3 & <50\\  
\ion{O}{vii} & 15& 5/4 & >25 &               &   &     &     & \ion{S }{vii} &  1& 2/3 & <50 & \ion{Ni}{iv}  & 38& 1/3 & <50\\  
\ion{O}{viii}&  1& 2/4 & >50 &               &   &     &     &               &   &     &     & \ion{Ni}{v}   & 48& 1/3 & <50\\
             &   &     &     &               &   &     &     &               &   &     &     & \ion{Ni}{vi}  &  1& 1/3 & <50\\
\hline
\end{tabular}
\tablefoot{Sources of atomic data: 1 -- TLUSTY; 2 -- Opacity and Iron Projects; 3 -- VALD;
4 -- NIST; 5 -- \citet{opc5}; 6 -- \citet{top11}; 7 -- \citet{top14}; 8 --
\citet{top9}; 9 -- \citet{pahole}; 10 -- \citet{bumez}; 11 -- \citet{napra};
12 -- \citet{naprane}; 13 -- \citet{top20}; 14 -- \citet{savej}; 15 -- Kurucz; 
16 -- \citet{zel1}, 17 -- \citet{zel5}, 18 -- \citet{zel4}, 19 -- \citet{zel3}.}
\end{table*}

The wind models of CSPNe were calculated using the global (unified) code METUJE
\citep{cmf1,cmfkont}. The global models provide self-consistent structure of the
stellar photosphere and radiatively-driven wind and, therefore, avoid artificial
splitting of these regions. The code solves the radiative transfer equation, the
kinetic (statistical) equilibrium equations, and hydrodynamic equations from a
nearly hydrostatic photosphere to supersonically expanding wind. The code
assumes stationary (time-independent) and spherically symmetric wind.

The radiative transfer equation was solved in the comoving frame (CMF) after
\citet{mikuh}. In the CMF radiative transfer equation, we account for line and
continuum transitions of elements given in Table~\ref{prvky}, which are relevant
in atmospheres of CSPNe. The inner boundary condition for the radiative transfer
equation was derived from the diffusion approximation.

The ionization and excitation state of considered elements (listed in
Table~\ref{prvky}) was calculated from the kinetic equilibrium equations, which are also
called the statistical equilibrium or NLTE equations. These equations account
for the radiative and collisional excitation, deexcitation, ionization
(including Auger ionization), and recombination. Part of the models of ions was
adopted from the TLUSTY model stellar atmosphere input data \citep[denoted as 1
in Table~\ref{prvky}]{ostar2003,bstar2006}. The remaining ionic models given in
Table~\ref{prvky} were prepared using the same strategy as in TLUSTY, that is, the
data are based on the Opacity and Iron Project calculations \citep{topt,zel0} or
they were downloaded from the NORAD
webpage\footnote{\url{http://norad.astronomy.ohio-state.edu}} and corrected with
the level energies and oscillator strengths available in the NIST database
\citep{nist}. The models of phosphorus ions were prepared using the data described
by \citet{pahole}. The models of ions are based on a superlevel concept as in
\citet{ostar2003,bstar2006}, that is, the low-lying levels are included
explicitly, while the higher levels are merged into superlevels. The bound-free
radiative rates in kinetic equilibrium equations are consistently calculated
from the CMF mean intensity. However, for the bound-bound rates, we used the
Sobolev approximation, which can be easily linearized for iterations.

We applied three different methods to solve the energy equation. In the
photosphere, we used differential and integral forms of the transfer equation
\citep{kubii}, while we applied the electron thermal balance method \citep{kpp} in
the wind. Individual terms in these equations are taken from the CMF radiative
field. The energy equations together with the remaining hydrodynamical equations,
that is, the continuity equation and the equation of motion, were solved
iteratively to obtain the wind density, velocity, and temperature structure.
These iterations were performed using the Newton-Raphson method. The models
include the radiative force due to line and continuum transitions calculated
from the CMF radiative field. However, to avoid possible negative velocity
gradients in the photosphere, we limited the continuum radiative force there to a
multiple (typically 2) of the radiative force due to the light scattering on
free electrons. The line data for the calculation of the line force (see
Table~\ref{prvky}) were taken from the VALD database (Piskunov et al.
\citeyear{vald1}, Kupka et al. \citeyear{vald2}) with some updates using the
NIST \citep{nist} and \citet{kur01} data. For \ion{Fe}{viii} -- \ion{Fe}{x,} we
used line data from the Kurucz
website\footnote{\url{http://kurucz.harvard.edu}}.

Our model predicts flow structure as a function of the base velocity, which
combined with the density gives the mass-loss rate. The code searches for such a
value of the mass-loss rate that allows the solution to smoothly pass through
the wind critical point, where the wind velocity is equal to the speed of
radiative-acoustic waves \citep{abbvln,thofe}. We calculated a series of wind
models with variable base velocity and searched for the base velocity, which
provides a smooth transonic solution with a maximum mass-loss rate using the
``shooting method'' \citep{cmfkont}. 

We used the TLUSTY plane-parallel static model stellar atmospheres
\citep{ostar2003,bstar2006} to derive the initial estimate of photospheric
structure and to compute the emergent fluxes for comparison purposes. The TLUSTY models
were calculated for the same effective temperature, surface gravity, and
chemical composition as the wind models.

\begin{figure}[t]
\centering
\resizebox{\hsize}{!}{\includegraphics{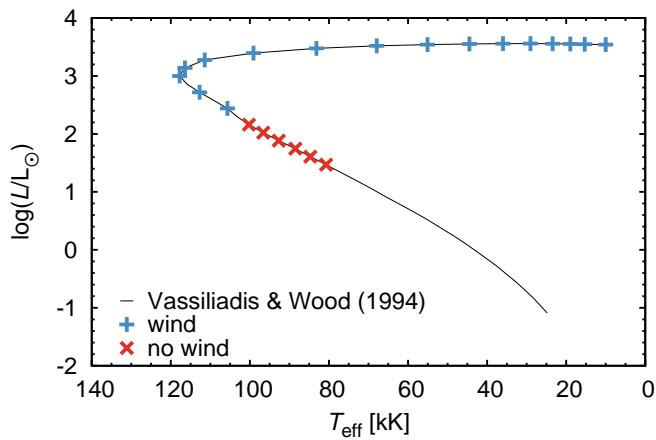}}
\caption{Adopted evolutionary track of the star with CSPN mass
$M=0.569\,M_\odot$ in HR diagram. Blue plus symbols and red crosses denote the
locations of studied models from Table~\ref{hvezpar} with and without wind.}
\label{chlazbt}
\end{figure}

\begin{table}
\caption{Adopted stellar parameters \citep[after][]{vasiwo} of the model grid
and predicted wind parameters for $M=0.569\,M_\odot$ and $Z=Z_\odot$.}
\centering
\label{hvezpar}
\begin{tabular}{r@{\hspace{2.0mm}}r@{\hspace{2.0mm}}r@{\hspace{2.0mm}}c@{\hspace{1.0mm}}c@{\hspace{0.5mm}}c@{\hspace{0mm}}r}
\hline
\hline
Model&\multicolumn{4}{c}{Stellar parameters} & \multicolumn{2}{c}{Prediction}\\
&\multicolumn{1}{c}{$\Teff$} & \multicolumn{1}{c}{$R_{*}$} & $\log g$ &
$\log(L/L_\odot)$ & $\dot M$ & \vnek \\
& \multicolumn{1}{c}{$[\text{K}]$} & \multicolumn{1}{c}{$[{R}_{\odot}]$} &
cgs & & [\msr] & [\kms] \\
\hline
T10\tablefootmark{$\ast$}\hspace{-5.2pt}  &  10000 & 19.6882 & 1.60 & 3.541 & $1.7\times10^{-10}$& 130\\
T15  &  15311 &  8.4665 & 2.34 & 3.548 & $1.0\times10^{-8}$ &  90\\
T18  &  18967 &  5.5425 & 2.71 & 3.552 & $2.0\times10^{-8}$ & 230\\
T23  &  23442 &  3.6451 & 3.07 & 3.556 & $3.8\times10^{-9}$ & 270\\
T29  &  29040 &  2.3834 & 3.44 & 3.559 & $2.4\times10^{-9}$ & 540\\
T35  &  35975 &  1.5513 & 3.81 & 3.558 & $3.8\times10^{-9}$ & 590\\
T44  &  44463 &  1.0097 & 4.19 & 3.553 & $6.0\times10^{-9}$ & 760\\
T55  &  55081 &  0.6497 & 4.57 & 3.542 & $4.7\times10^{-9}$ &1150\\
T67  &  67920 &  0.4166 & 4.95 & 3.520 & $3.0\times10^{-9}$ &1550\\
T83  &  83176 &  0.2641 & 5.35 & 3.476 & $2.9\times10^{-9}$ &2630\\
T99  &  99083 &  0.1695 & 5.74 & 3.395 & $1.3\times10^{-9}$ &2530\\
T111 & 111429 &  0.1170 & 6.06 & 3.277 & $1.2\times10^{-10}$&3870\\
T116 & 116413 &  0.0915 & 6.27 & 3.139 & $6.1\times10^{-10}$&2330\\
T117 & 117761 &  0.0761 & 6.43 & 2.999 & $2.6\times10^{-10}$&2170\\
T112 & 112720 &  0.0601 & 6.64 & 2.718 & $5.4\times10^{-11}$&1830\\
T105\tablefootmark{$\ast$}\hspace{-5.2pt} & 105682 &  0.0496 & 6.80 & 2.440 & $5.1\times10^{-13}$&1890\\
T100 & 100231 &  0.0400 & 6.99 & 2.160 & \multicolumn{2}{c}{no wind}\\
T96  &  96605 &  0.0367 & 7.07 & 2.021 & \multicolumn{2}{c}{no wind}\\
T92  &  92683 &  0.0340 & 7.13 & 1.883 & \multicolumn{2}{c}{no wind}\\
T88  &  88512 &  0.0318 & 7.19 & 1.745 & \multicolumn{2}{c}{no wind}\\
T84  &  84723 &  0.0296 & 7.25 & 1.607 & \multicolumn{2}{c}{no wind}\\
T80  &  80724 &  0.0278 & 7.30 & 1.470 & \multicolumn{2}{c}{no wind}\\
\hline
\end{tabular}
\tablefoot{\tablefoottext{$\ast$}{Core-halo model.}}
\end{table}

The wind models were calculated with stellar parameters (given in
Table~\ref{hvezpar}) taken from the post-AGB evolutionary tracks of \citet[see
also Fig.~\ref{chlazbt}]{vasiwo}. We selected the track with an initial
(main-sequence) mass of $1\,M_\odot$, a helium abundance of 0.25, and a metallicity of
0.016, which yields a CSPN mass of $M=0.569\,M_\odot$. This corresponds to a typical
mass of CSPN and white dwarfs, which is about $0.55-0.65\,M_\odot$
\citep[e.g.,][]{koest,stasihmot,kuplan,nenijoris,sokol}. Although some deviations
of the chemical compositions of CSPN atmospheres from the solar chemical
composition can be expected as a result of stellar evolution, most CSPNe and
their nebulae have near-solar chemical composition
\citep[e.g.,][]{stasiz,wali,militka}. Therefore, we assumed solar abundances
$Z_\odot$ according to \citet{asp09}. 

To get a better convergence, our code assumes an inner boundary radius that is fixed deep
in the photosphere. As a result of this, for a given stellar luminosity, the
conventionally defined stellar radius is slightly higher and the effective
temperature is slightly lower than what is assumed here. This is not a problem for
the stars with a higher gravity (with thin atmospheres), but this may have a larger
impact on atmospheric models of the stars with lower effective temperatures,
which have more extended atmospheres.

Our code finds the wind critical point, which appears slightly above the
photosphere. There the mass-loss rate of our models is determined. In some
cases, when the initial range of radii was too narrow, the code was unable to
find the critical point. A simple solution would be to extend the subcritical
model to include the region of the critical point. In some cases, this simple
approach failed during subsequent iterations, even if another calculation with
a different initial estimate of the solution was able to find a physical
solution. This is most likely caused by a complex wind structure that is close to the
photosphere, where the temperature bump exists, which significantly influences
the density distribution of  the subsonic part of the wind. A similar problem can
also appear in the codes that use direct integration of hydrodynamical
equations.

For some stellar parameters (denoted using an asterisk in Table~\ref{hvezpar}),
we were unable to get converged global models. For these models, we calculated
core-halo models with TLUSTY model atmosphere emergent flux as a boundary
condition \citep{cmf1}. This may lead to a slight overestimation of the
mass-loss rate \citep{cmfkont} for these particular models.

\begin{table*}[t]
\caption{Parameters of the fit of the mass-loss rate in Eq.~\eqref{dmdtcspn}.}
\label{tabfitl}
\begin{tabular}{*{15}{c}}
\hline
\hline
$\alpha$ & $\zeta$ & $Z_\odot$ &
$m_1$ & $m_2$ & $m_3$ & $m_4$ & $T_1$ & $T_2$ & $T_3$ & $T_4$ &
$\Delta T_1$ & $\Delta T_2$ & $\Delta T_3$ & $\Delta T_4$\\
&& & \multicolumn{4}{c}{[$10^{-9}\,\msr$]} & \multicolumn{4}{c}{[kK]} &
\multicolumn{4}{c}{[kK]}\\
\hline
1.63 & 0.53 & 0.0134 &
2.51 & 0.72 & 0.48 & 0.33 & 18.3 & 46.4 & 84.6 & 116.7 & 3.63 & 17.7 & 20.4 &
1.7\\
\hline
\end{tabular}
\centering
\end{table*}

\section{Calculated wind models}

Parameters of calculated wind models (mass-loss rates $\dot M$ and terminal
velocities $v_\infty$) are given in Table~\ref{hvezpar}. The wind mass-loss rate
mostly depends on the stellar luminosity \citep[e.g.,][]
{cak,vikolamet,powrdyn,cmfkont}; consequently, the mass-loss rate is nearly
constant (on the order of $10^{-9}\,\msr$) during the evolution of a star from the
AGB to the top of the white dwarf cooling track. The mass-loss rate increases with
temperature by more than an order of magnitude for the coolest stars at the
beginning of the adopted evolutionary tracks. This is caused by the increase in
the UV flux for frequencies that are higher than that of the Lyman jump where most of the
strong resonance lines, which drive the wind, appear. When the star settles on the
white dwarf cooling track, its luminosity starts to decrease leading to the
decrease in the mass-loss rate. We have not found any winds for stars at the
white dwarf cooling track after reaching effective temperatures lower than about
100\,kK (Fig.~\ref{chlazbt}).

\begin{figure}[t]
\centering
\resizebox{\hsize}{!}{\includegraphics{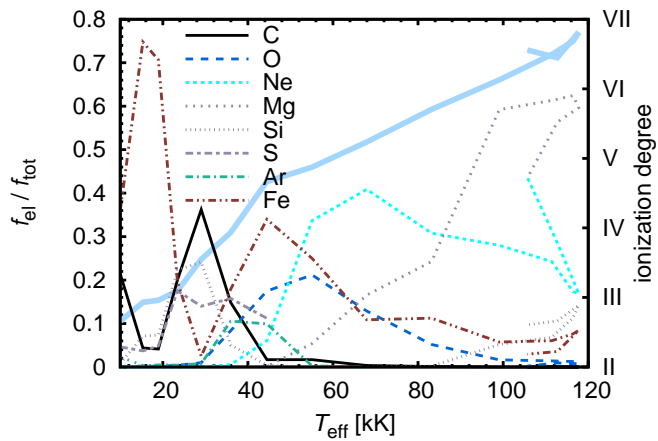}}
\caption{Relative contribution of individual elements to the radiative force at
the critical point as a function of the effective temperature. Thick blue line
denotes the mean ionization degree that contributes to the radiative force,
$\sum f_iz_i/(\sum f_i)$, where $z_i$ is the charge of ion $i$ and $f_i$ is its
contribution to the radiative force.}
\label{btsil}
\end{figure}

Fig.~\ref{btsil} shows the relative contribution of individual elements to the
radiative force at the critical point. From Fig.~\ref{btsil}, it follows that
winds of CSPN are mostly accelerated by C, O, Ne, Mg, and Fe. The contribution of
individual elements to the radiative force strongly varies with effective
temperature due to changes in the ionization structure. This is demonstrated in
the plot of the mean ionization degree that mostly contributes to the radiative
force (Fig.~\ref{btsil}). The degree changes from\,\ion{}{iii} for the coolest
stars to\,\ion{}{vii} for the hottest stars. The wind mass-loss rate is
relatively high around $T_\text{eff}\approx20\,$kK; consequently, iron lines are
optically thick and dominate the radiative driving \citep{vikolamet}. With
an increasing temperature, iron ionizes from \ion{Fe}{iii} to \ion{Fe}{iv}, which
accelerates the wind less efficiently. Consequently, lighter elements, such as
carbon, oxygen, and silicon, take over the wind acceleration. However, for even
higher temperatures, these elements appear in closed shell configurations as a
He-like ion (carbon) and a Ne-like ion (silicon). These ions have resonance
lines around 10\,\AA, where the stellar flux is low, and as a result they do not
significantly accelerate the wind. On the other hand, neon and magnesium, with
abundances that are higher than silicon, appear in open shell configurations with a larger
number of lines that are capable of driving the wind at higher temperatures.

The wind mass-loss rate mostly depends on the stellar luminosity, which is constant
during the post-AGB evolution. As a result, Table~\ref{hvezpar} mostly shows
just mild variations of the mass-loss rate with temperature, connected with
changes of contribution of individual ions to the radiative force during the
evolution (Fig.~\ref{btsil}). Models around $\Teff=20\,$kK most significantly
deviate from this behavior. The mass-loss rate from the model T18 to the model
T29 decreases by a factor of about 8. This is caused by the decrease in the
contribution of iron to the radiative driving (see Fig.~\ref{btsil}). Iron
around $\Teff=20\,$kK ionizes from \ion{Fe}{iii} to \ion{Fe}{iv}, which
accelerates the wind less efficiently. This is accompanied by the increase in
the terminal velocity. The behavior of the mass-loss rate and of the terminal
velocity corresponds to the bistability jump around $T_\text{eff}\approx21\,$kK, which is
found in wind models of B supergiants \citep{bista, vikolabis, vinbisja}.
However, the observational behavior of the mass-loss rates of B supergiants
around the jump does not seem to follow the theoretical expectations
\citep{vysbeta,kuraci,hau}. The reason for different behavior of
observationally determined mass-loss rates is unclear and is likely caused by
inadequate wind models used to determine the mass-loss rates either from
observations or from theory.

\begin{figure*}[t]
\centering \resizebox{0.33\hsize}{!}{\includegraphics{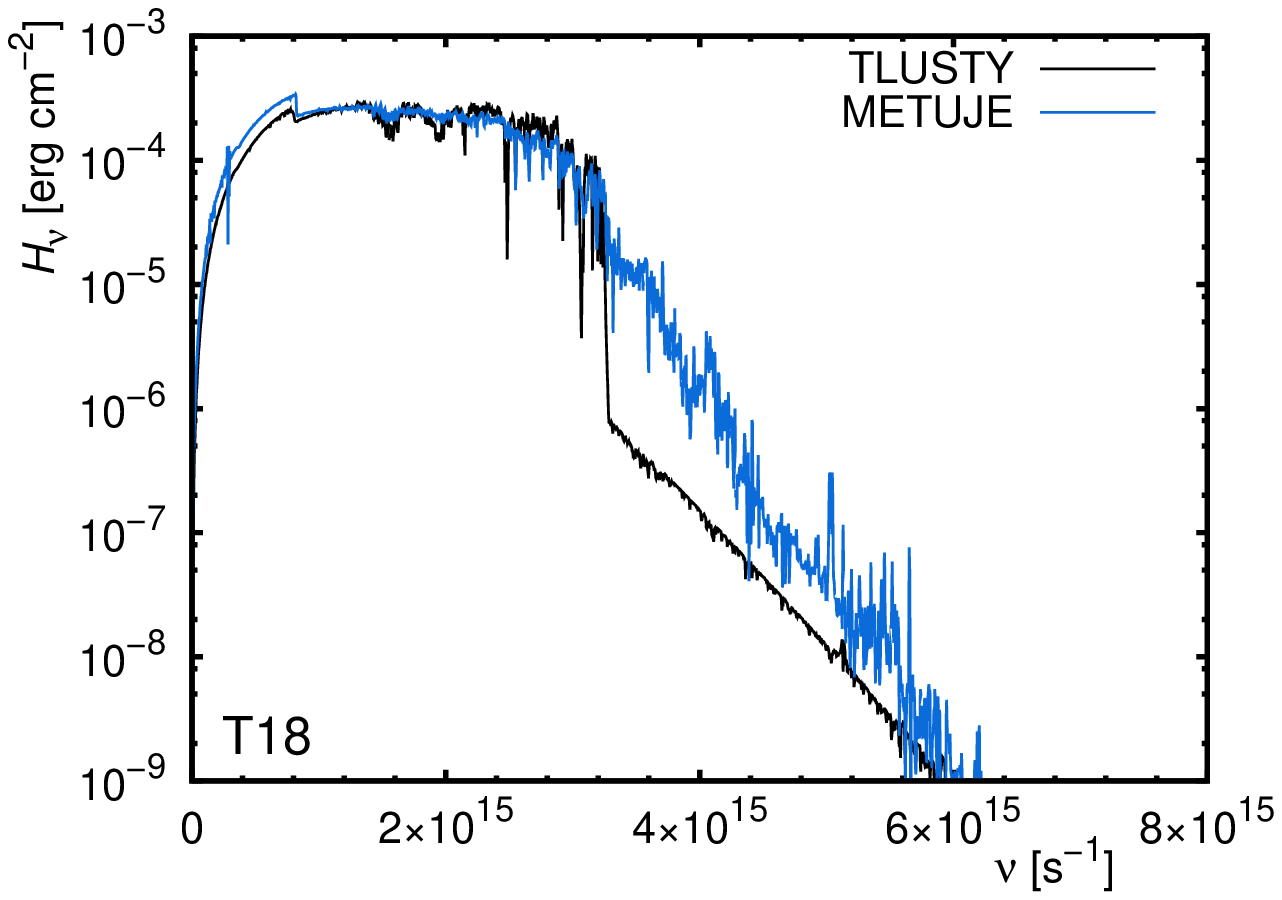}}
\centering \resizebox{0.33\hsize}{!}{\includegraphics{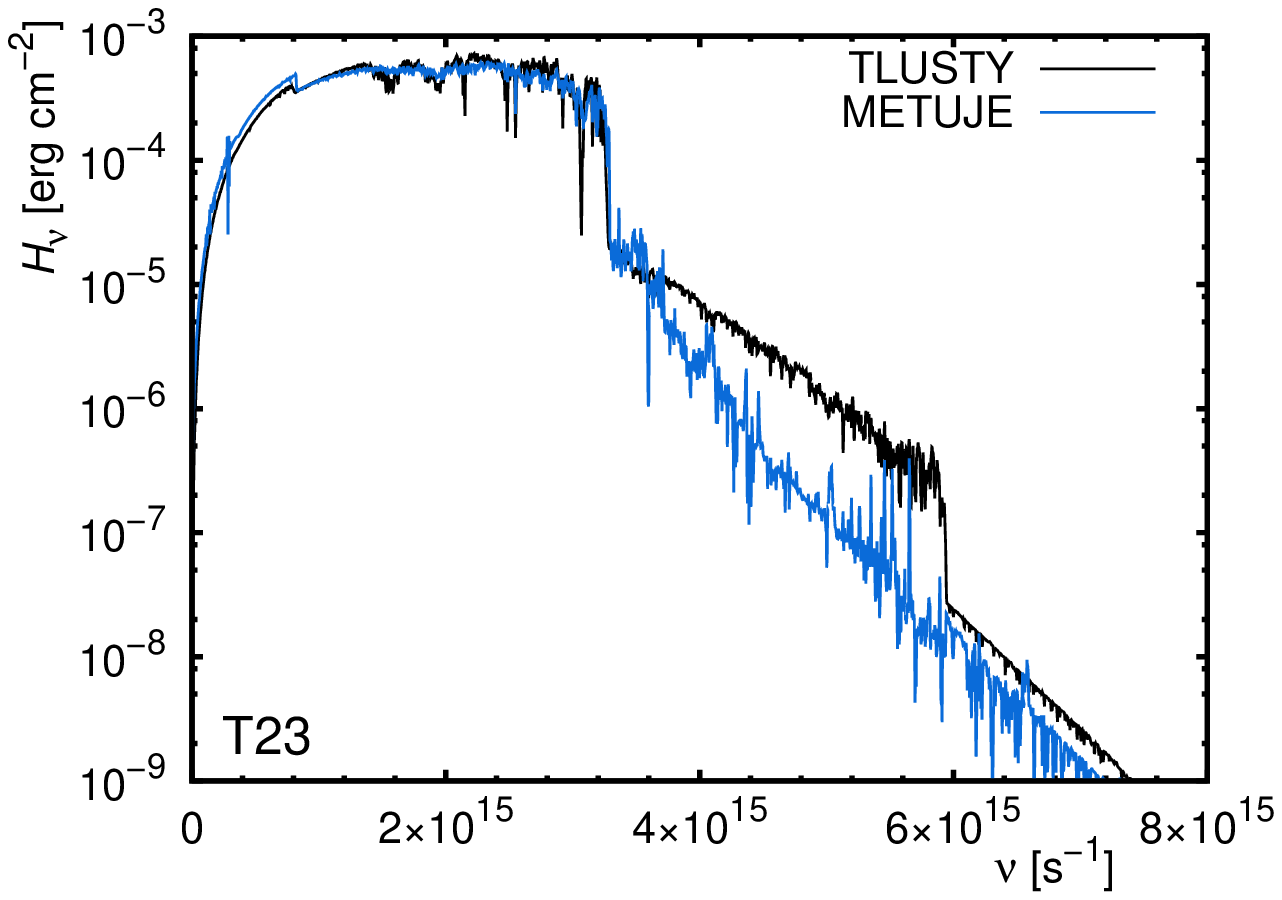}}
\centering \resizebox{0.33\hsize}{!}{\includegraphics{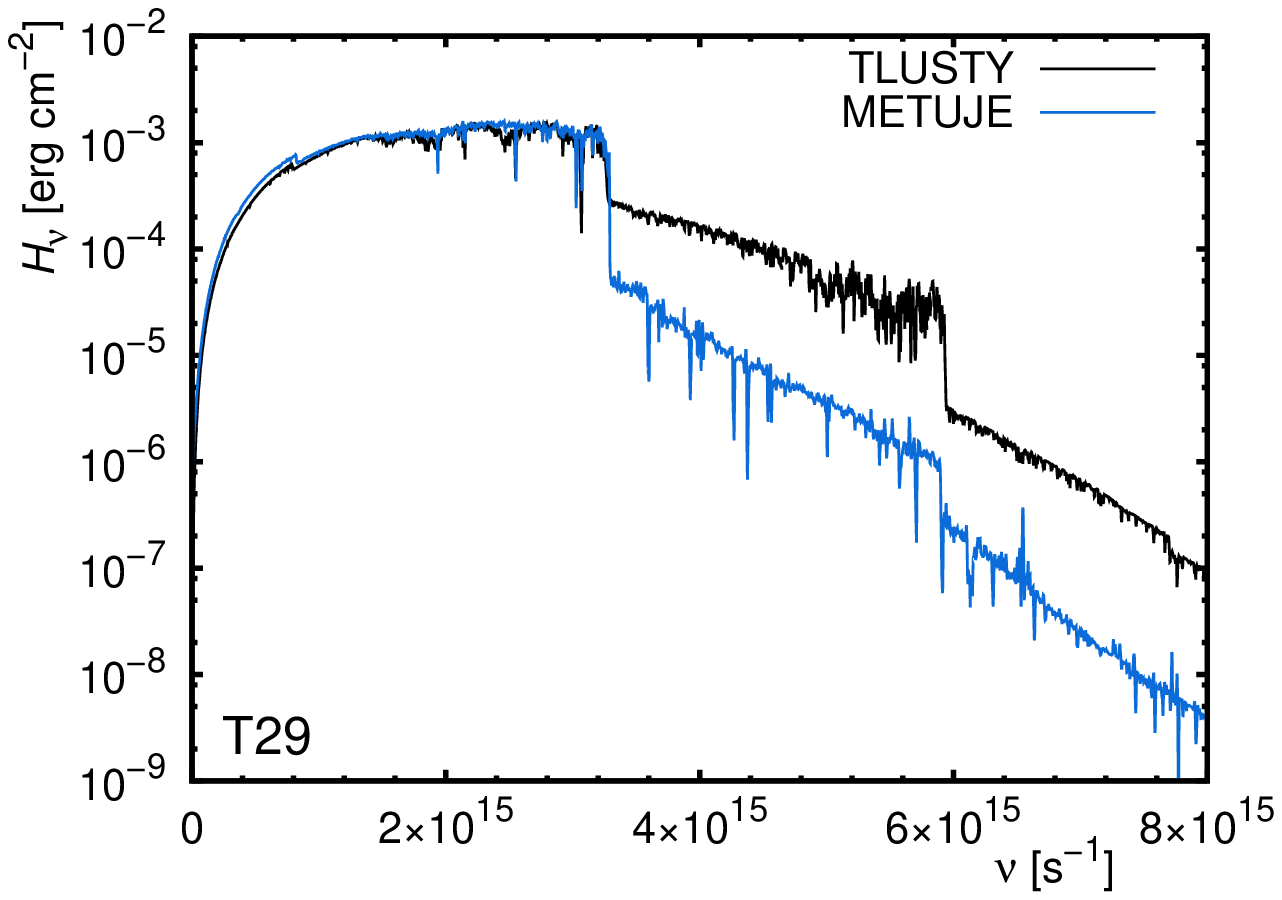}}
\centering \resizebox{0.33\hsize}{!}{\includegraphics{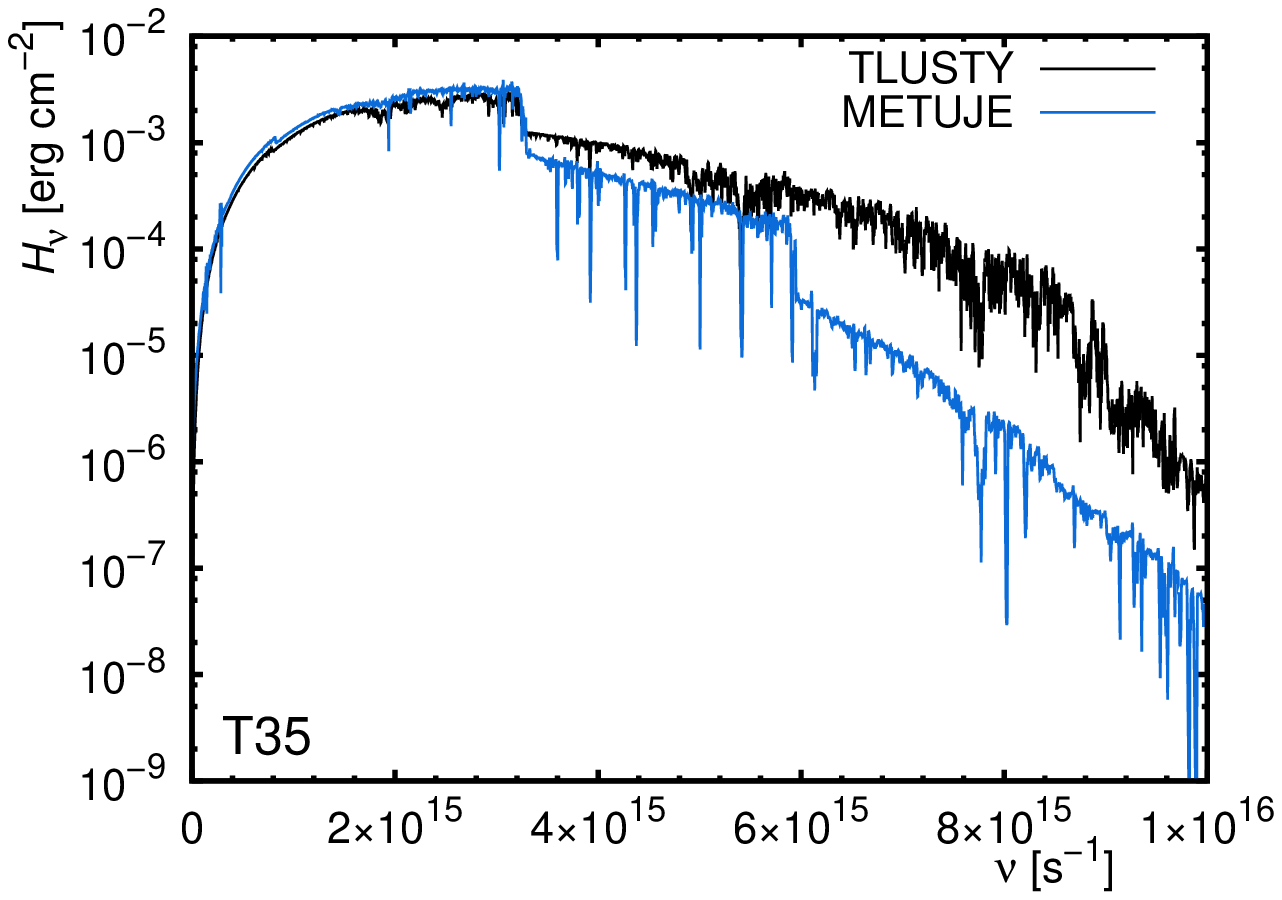}}
\centering \resizebox{0.33\hsize}{!}{\includegraphics{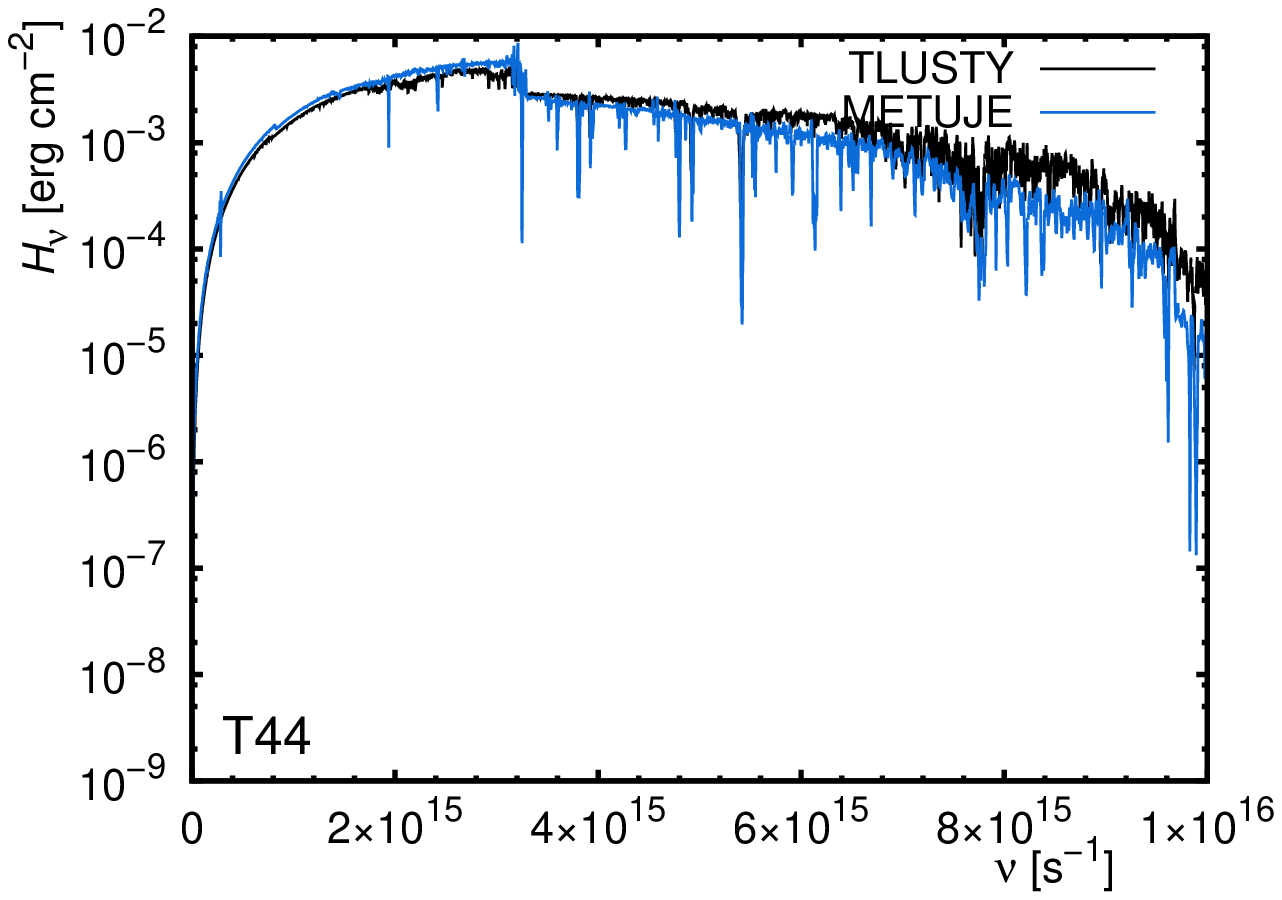}}
\centering \resizebox{0.33\hsize}{!}{\includegraphics{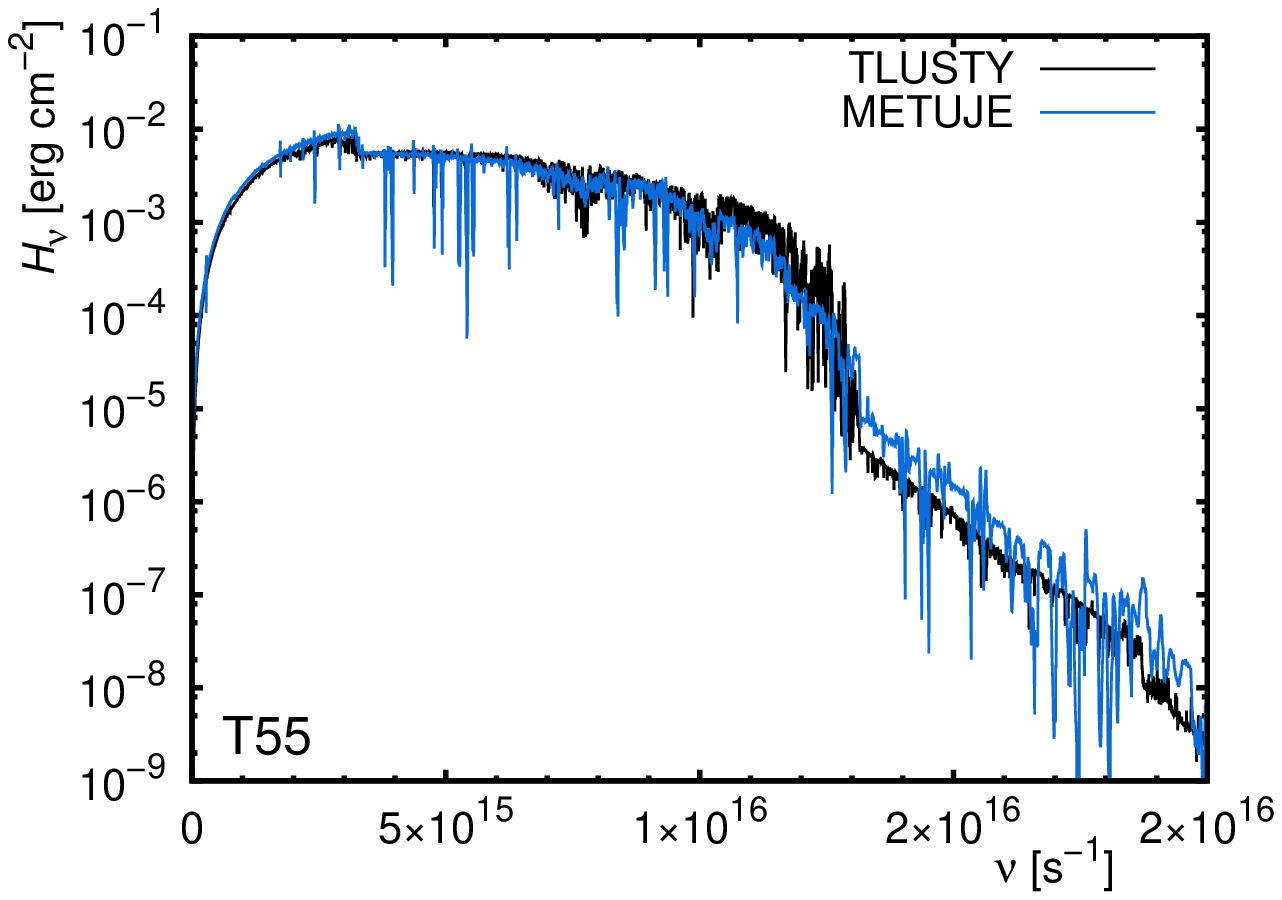}}
\centering \resizebox{0.33\hsize}{!}{\includegraphics{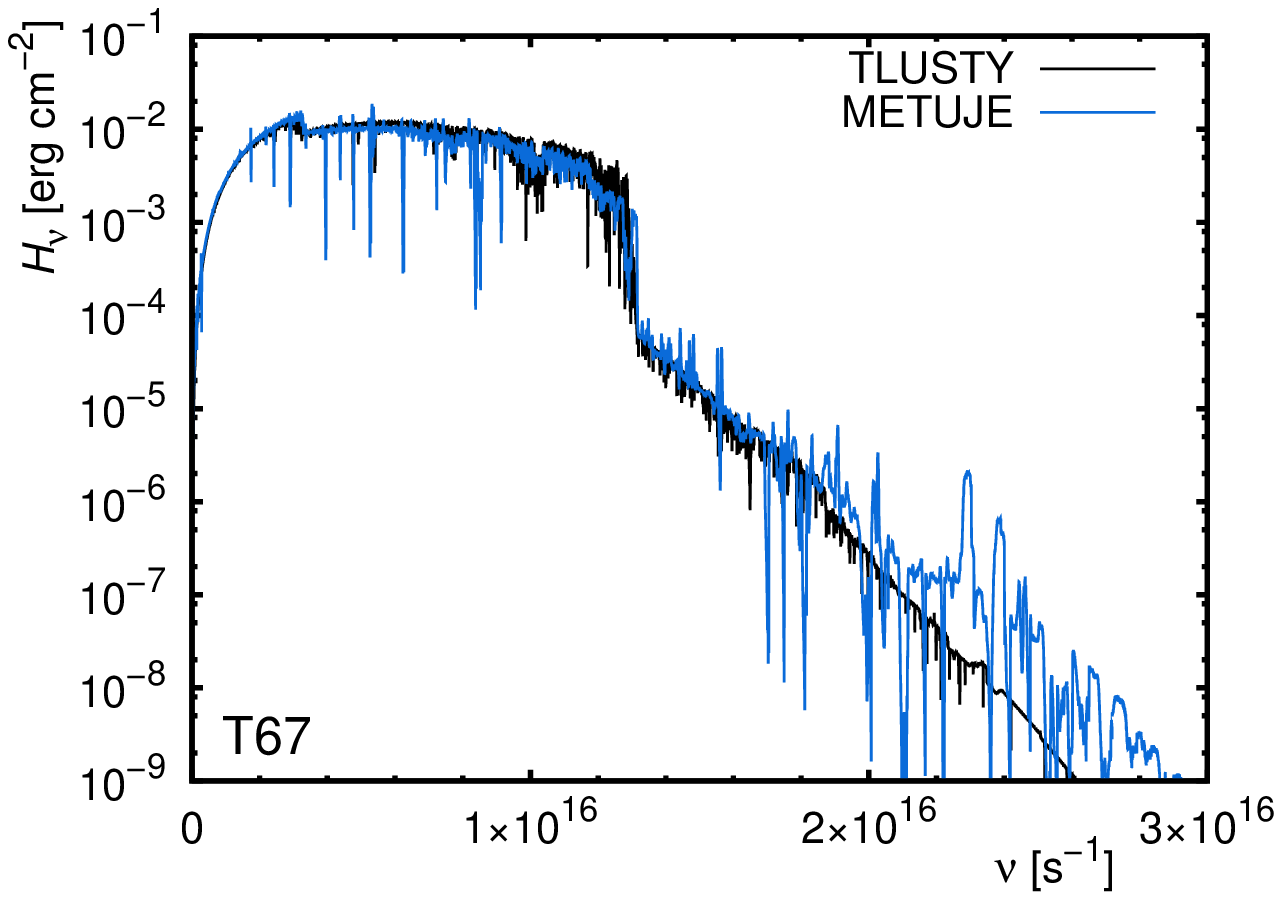}}
\centering \resizebox{0.33\hsize}{!}{\includegraphics{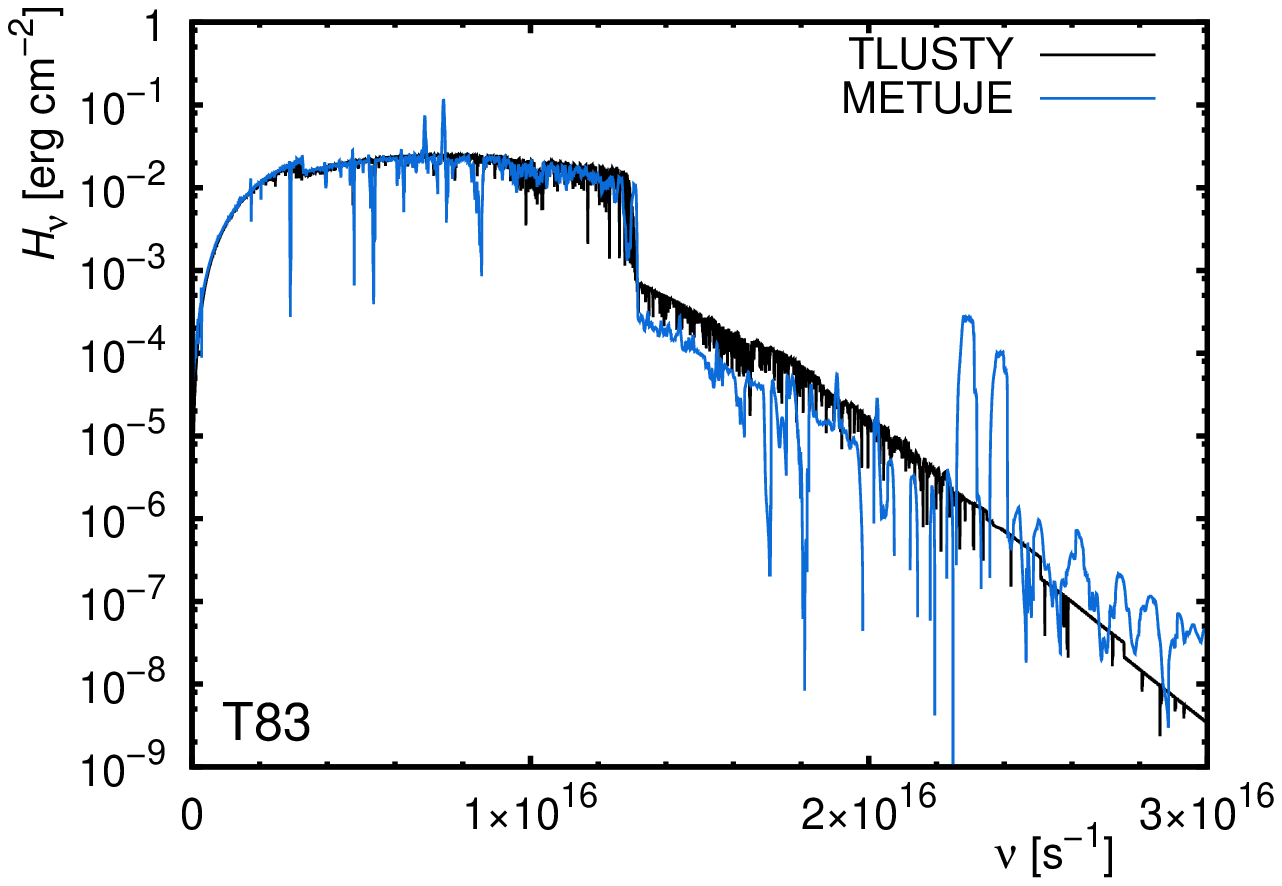}}
\centering \resizebox{0.33\hsize}{!}{\includegraphics{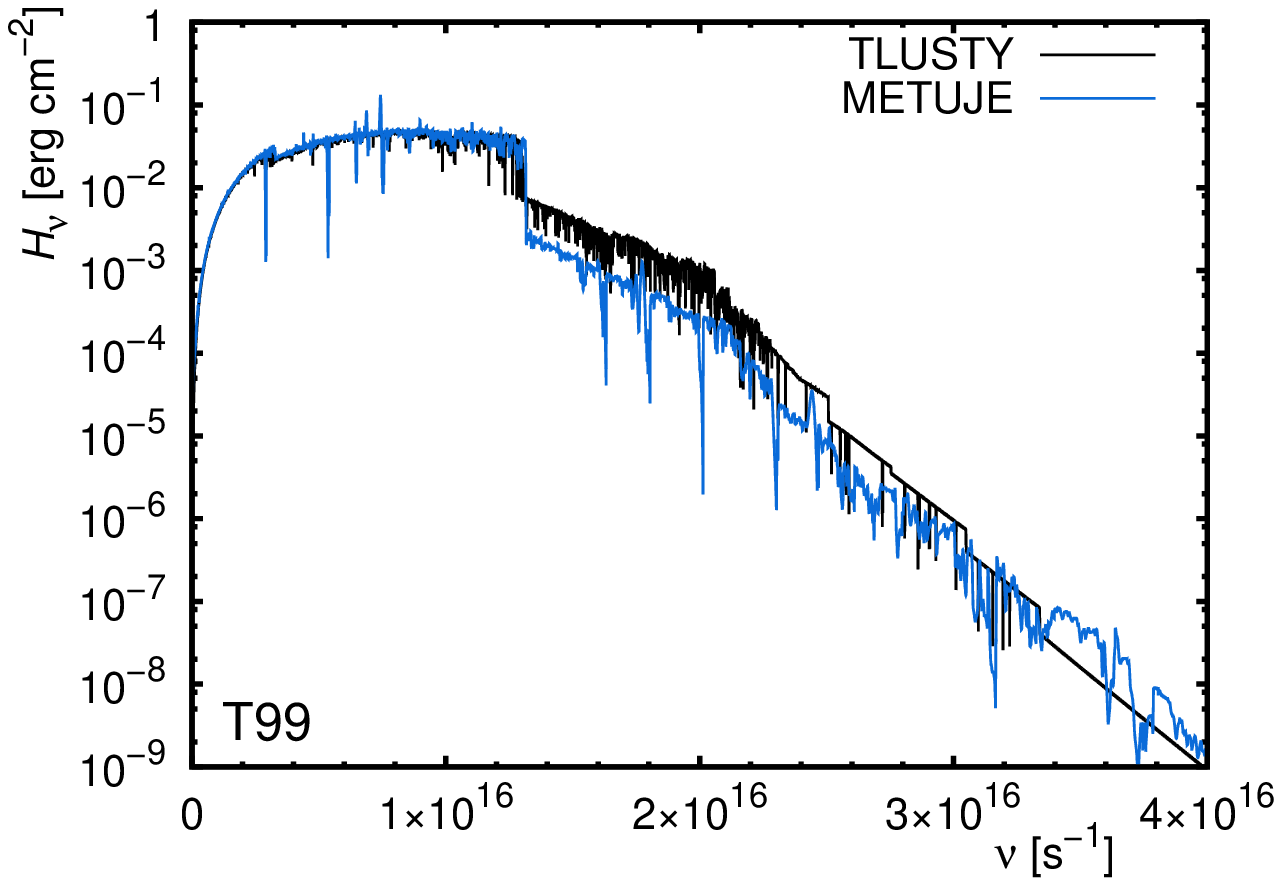}}
\centering \resizebox{0.33\hsize}{!}{\includegraphics{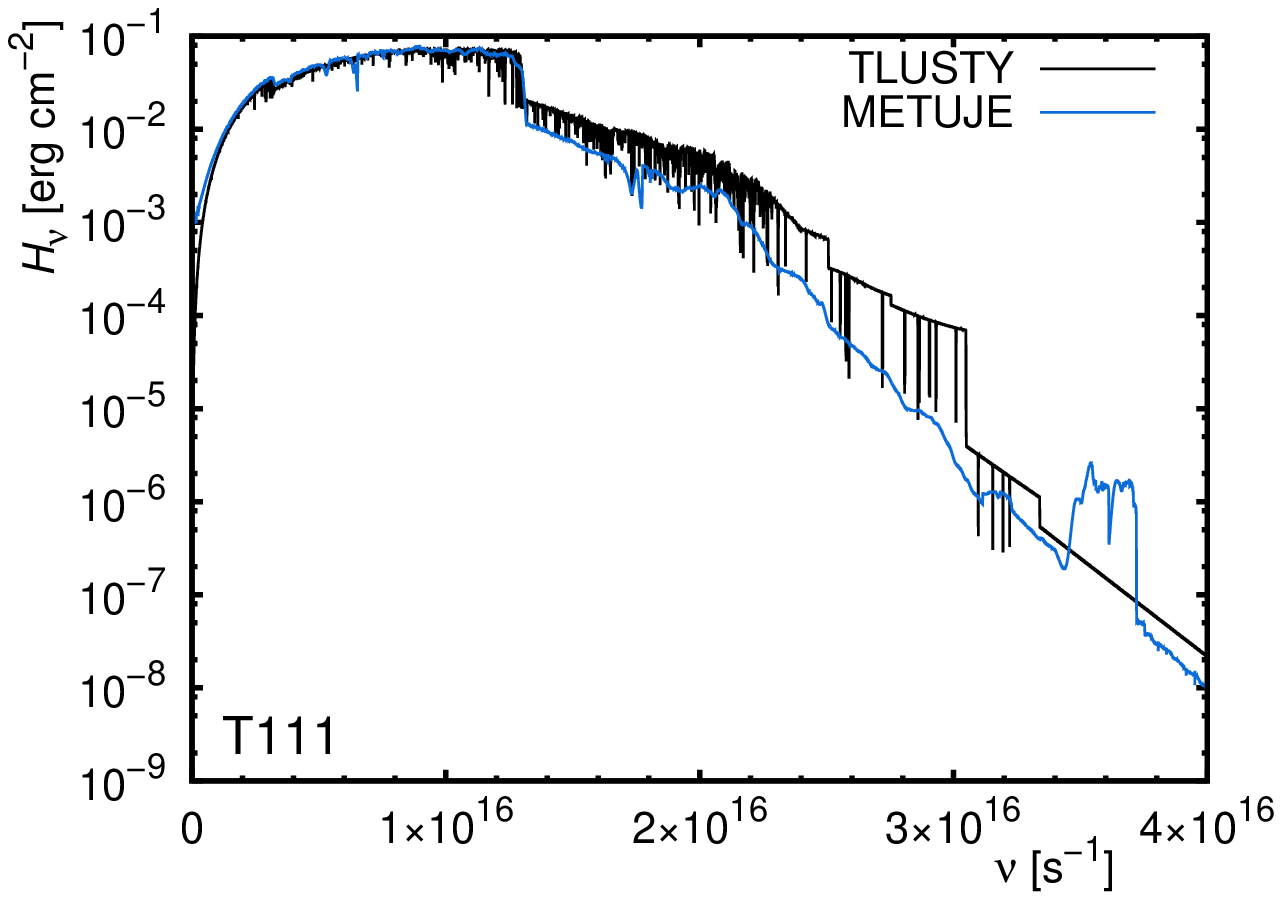}}
\centering \resizebox{0.33\hsize}{!}{\includegraphics{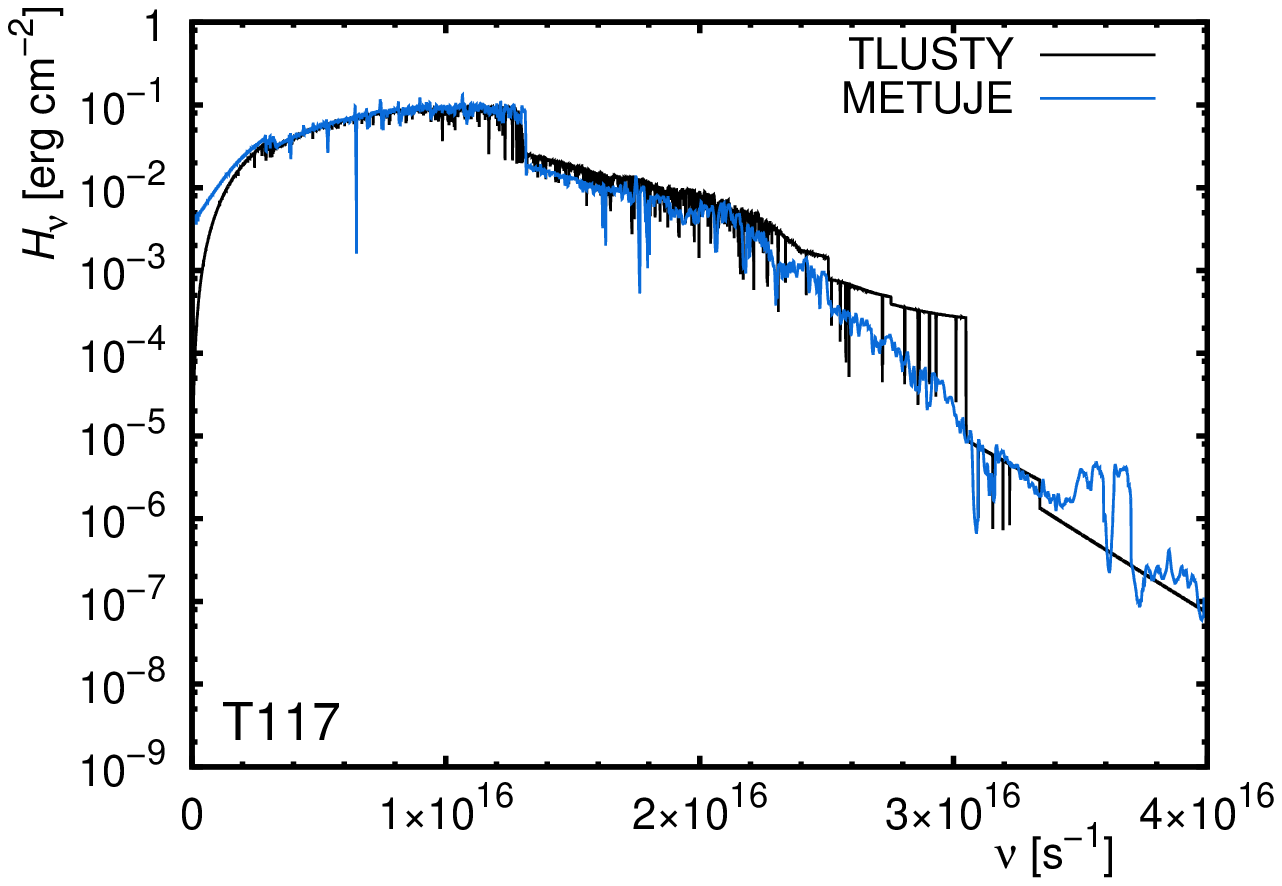}}
\centering \resizebox{0.33\hsize}{!}{\includegraphics{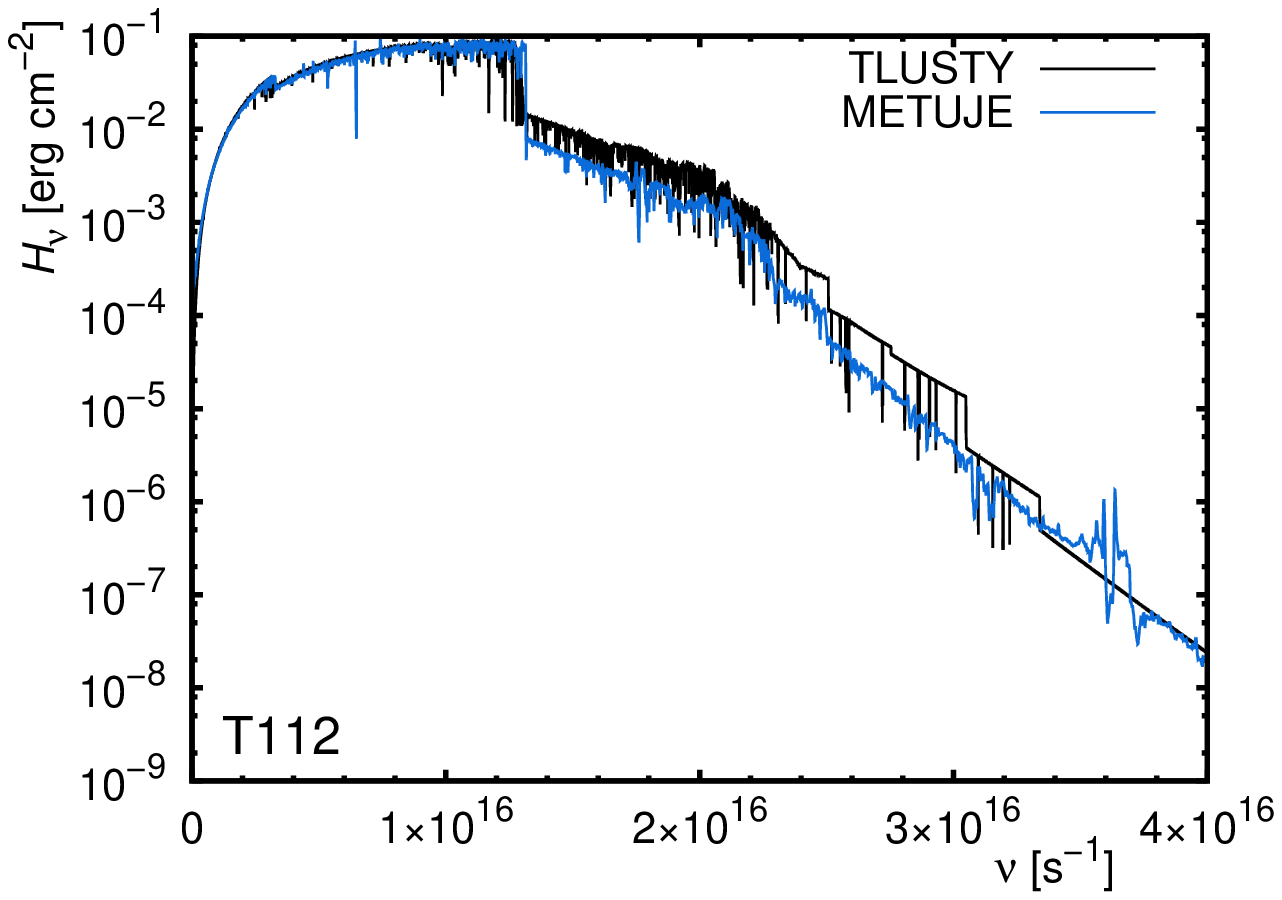}}
\caption{Comparison of the emergent flux from TLUSTY (black line) and METUJE
(blue line) models smoothed by a Gaussian filter. The graphs are plotted for
individual models from Table~\ref{hvezpar} (denoted in the plots).}
\label{bttok}
\end{figure*}

There is another broad and weak maximum of the mass-loss rates around
$\Teff=45\,$kK, which was also found in the models of \citet{btpau}. The maximum is
caused by a higher radiative flux in the far-UV domain (cf.~the flux
distributions of T35 and T44 models in Fig.~\ref{bttok}), which leads to an
increase in the radiative force due to \ion{Fe}{v} (Fig.~\ref{btsil}).

Another case that deviates from strict $\dot M\sim L$ dependence is the T116
model. Although the corresponding star has a slightly lower luminosity than that
of the T111 model, the mass-loss rate of T116 is higher than the mass-loss rate of
the T111 model (see Table~\ref{hvezpar}). This is caused by the change in the
ionization of magnesium that most significantly contributes to the radiative
force at these temperatures (see Fig.~\ref{btsil}). In most cases, higher
ionization leads to a weaker radiative force, but \ion{Mg}{vii} has more
resonance lines that are close to the flux maximum than \ion{Mg}{vi}; consequently, as a
result of higher ionization, the radiative force and the mass-loss rate increase
in this case. The increase in the mass-loss rate is accompanied by the decrease
in the terminal velocity (Table~\ref{hvezpar}). This behavior of the mass-loss
rate and the terminal velocity resembles the bistability jump around
$T_\text{eff}\approx20\,$kK, which is due to the change in iron ionization
\citep{bista,vinbisja}.

The wind mass-loss rate of the calculated models on the leftward (heating) part
of the evolutionary track can be fit as a function of the effective
temperature with a sum of four Gaussians
\begin{equation}
\label{dmdtcspn}
\dot M(T_\text{eff},L,Z)=\zav{\frac{L}{10^3\,L_\odot}}^{\alpha}
  \zav{\frac{Z}{Z_\odot}}^{\zeta}
\sum_{i=1}^4 m_i \exp\hzav{-\zav{\frac{T_\text{eff}-T_i}{\Delta T_i}}^2},
\end{equation}
where $m_i$, $T_i$, and $\Delta T_i$ are parameters of the fit given in
Table~\ref{tabfitl}. Eq.~\eqref{dmdtcspn} fits the numerical results with
an average precision of about 10\%. To account for the dependence of the wind mass-loss
rate on the stellar luminosity, we introduced the additional parameter $\alpha$ in
Eq.~\eqref{dmdtcspn}, which was kept fixed during fitting. We adopted its value
from O star wind models \citep[Eq.~(11)]{cmfkont}. The metallicity dependence
parameter $\zeta$ was derived by fitting results of nonsolar metallicity models
(see Sect.~\ref{inmet}), assuming a solar mass fraction of heavy elements $Z_\odot$
after \citet{asp09}. Eq.~\eqref{dmdtcspn} can only be used for the stars within
the studied temperature range, that is, for $T_\text{eff}=10-117\,$kK.

Wind terminal velocity is proportional to the surface escape speed
\citep[e.g.,][]{cak,lsl,vysbeta}. The stellar luminosity is constant after
leaving the AGB phase, but the effective temperature increases and, therefore, the
stellar radius decreases. As a result, the surface escape speed rises
and, in addition, the wind terminal velocity increases from $100\,\kms$ to about $ 4
000\,\kms$. The wind driving becomes less efficient when the star approaches the
white dwarf cooling track; therefore, with a decreasing mass-loss rate, the
wind terminal velocity also decreases.

\begin{table}[t]
\caption{Calculated number of ionizing photons per unit of surface area 
$\log \zav{Q/1\,\text{cm}^{-2}\,\text{s}^{-1}}$.}
\label{pocq}
\begin{tabular}{lcccccc}
\hline\hline
Model & \multicolumn{3}{c}{TLUSTY} & \multicolumn{3}{c}{METUJE}\\
& \ion{H}{i} & \ion{He}{i}& \ion{He}{ii}
& \ion{H}{i} & \ion{He}{i}& \ion{He}{ii}\\
\hline
T15  & 18.93 & 14.20 &  1.31 & 20.65 & 14.42 & $-0.07$ \\
T18  & 20.22 & 16.17 &  5.47 & 21.53 & 17.13 &  1.16 \\
T23  & 21.79 & 18.55 &  9.26 & 21.66 & 18.30 &  4.98 \\
T29  & 23.12 & 20.76 & 12.51 & 22.19 & 19.55 &  9.89 \\
T35  & 23.97 & 23.02 & 16.05 & 23.65 & 21.85 & 12.90 \\
T44  & 24.53 & 23.91 & 18.93 & 24.40 & 23.66 & 16.70 \\
T55  & 24.97 & 24.54 & 20.73 & 24.92 & 24.47 & 21.03 \\
T67  & 25.36 & 25.05 & 22.04 & 25.30 & 24.96 & 22.10 \\
T83  & 25.71 & 25.48 & 23.22 & 25.68 & 25.44 & 22.87 \\
T99  & 25.99 & 25.81 & 24.41 & 26.00 & 25.81 & 23.94 \\
T111 & 26.18 & 26.02 & 24.98 & 26.18 & 26.01 & 24.69 \\
T116 & 26.25 & 26.10 & 25.12 & 26.25 & 26.10 & 24.93 \\
T117 & 26.27 & 26.12 & 25.12 & 26.27 & 26.13 & 24.97 \\
T112 & 26.21 & 26.05 & 24.80 & 26.19 & 26.04 & 24.54 \\
\hline
\end{tabular}
\end{table}

In Fig.~\ref{bttok} we compare emergent fluxes from our global models with
fluxes derived from plane-parallel TLUSTY photosphere models. There is relatively good agreement between these fluxes for frequencies below the Lyman
jump. However, for $\Teff\lesssim40\,$kK, the global models predict significantly
lower fluxes in the region of Lyman continuum due to the blocking of radiation
by the stellar wind and due to deviations from spherical symmetry \citep{kuhumi,
scspn}. These trends are also apparent in Table~\ref{pocq} where we give the
number of ionizing photons per unit of surface area
\begin{equation}
Q=4\pi\int_{\nu_0}^\infty \frac{H_\nu}{h\nu}\,\de\nu,
\end{equation}
where $H_\nu$ is the Eddington flux and the ionization frequency $\nu_0$
corresponds to ionization frequencies of \ion{H}{i}, \ion{He}{i}, and
\ion{He}{ii}. From the table, it follows that the number of \ion{H}{i} ionizing
photons can be derived from plane-parallel photospheric models for
$\Teff\gtrsim40\,$kK and that the plane-parallel models predict a reliable number
of \ion{He}{i} ionizing photons for $\Teff\gtrsim50\,$kK. On the other hand,
wind absorption above \ion{the He}{ii} ionization jump is so strong that the
plane-parallel models never predict the correct number of \ion{He}{ii} ionizing
photons for any star with wind. However, even global models are not able to
predict \ion{He}{ii} ionizing flux reliably because there are even significant
differences in predicting the \ion{He}{ii} ionizing flux between
individual global models \citep{fastpuls}.

Our derived number of \ion{H}{i} and \ion{He}{i} ionizing photons for the models
T35 and T44 agree typically within 0.1 -- 0.2~dex with the results for
dwarfs and supergiants that are published elsewhere \citep{moksit,fastpuls,okali}. However, the
differences in \ion{He}{ii} ionizing photons \citep{pahole} are more significant
and reflect involved model assumptions.

\begin{table}[t]
\caption{Wavelengths (in \AA) of the most prominent metallic wind lines in
individual models.}
\label{vitcar}
\begin{tabular}{ll}
\hline
T15  & \ion{Si}{iii} 1295, 1297, 1299, 1301, 1303;\\
     & \ion{C}{ii} 1335, 1336; \ion{Si}{iv} 1394, 1403;
       \ion{Al}{iii} 1855, 1863; \\
T18  & \ion{C}{iii} 977; \ion{N}{iii} 990, 992; \ion{S}{iii} 1012, 1016, 1021;\\
     & \ion{C}{ii} 1036, 1037; \ion{N}{ii} 1084, 1085, 1086;\\
     & \ion{Si}{iii} 1108, 1110, 1113; \ion{Si}{iv} 1122, 1128;
       \ion{C}{iii} 1176;\\ &\ion{Si}{iii} 1207, 1295, 1297, 1299, 1301, 1303;\\
     & \ion{C}{ii} 1335, 1336; \ion{Si}{iv} 1394, 1403;
       \ion{Al}{iii} 1855, 1863; \\
T23  & \ion{C}{iii} 977; \ion{N}{iii} 990, 992; \ion{C}{iii} 1176;
       \ion{Si}{iii} 1207; \\ & \ion{Si}{iv} 1394, 1403;
       \ion{C}{iv} 1548, 1551; \ion{Al}{iii} 1855, 1863; \\ &
       \ion{Na}{i} 5890, 5896\\
T29  & \ion{C}{iii} 977; \ion{N}{iii} 990, 992; \ion{C}{iii} 1176;
       \ion{Si}{iv} 1394, 1403; \\ & 
       \ion{C}{iv} 1548, 1551  \\
T35  & \ion{O}{iii} 702, 703, 704; \ion{O}{iv} 788, 790; \\
     & \ion{N}{iv} 922, 923, 924; \ion{C}{iii} 977; \ion{N}{iii} 990, 992;\\ 
     & \ion{C}{iii} 1176; \ion{Si}{iv} 1394, 1403; \ion{C}{iv} 1548, 1551\\
T44  & \ion{O}{iv} 553, 554, 555; \ion{O}{iii} 702, 703, 704; \\
     & \ion{O}{v} 759, 760, 761, 762; \ion{O}{iv} 788, 790; \\
     & \ion{N}{iv} 922, 923, 924; \ion{S}{vi} 933, 945; \ion{N}{v} 1239, 1243;\\
     & \ion{C}{iv} 1548, 1551\\
T55  & \ion{Ne}{iv} 470; \ion{Ne}{v} 480, 481, 483; \ion{O}{iv} 553, 554, 555;\\
     & \ion{Ne}{v} 570, 572; \ion{O}{iv} 608, 610, 625, 626; \\
     & \ion{O}{v} 630, 759, 760, 761, 762; \ion{O}{iv} 788, 790;\\
     & \ion{O}{vi} 1032, 1038; \ion{N}{v} 1239, 1243; \ion{C}{iv} 1548, 1551\\
T67  & \ion{Mg}{v} 351, 352, 353, 354, 355; \ion{Ne}{v} 358, 359;\\
     & \ion{Ne}{v} 480, 481, 483; \ion{Ne}{vi} 559, 563; \ion{Ne}{v} 570, 572;\\
     & \ion{O}{v} 630, 759, 760, 761, 762; \ion{O}{vi} 1032, 1038; \\
     & \ion{N}{v} 1239, 1243; \ion{O}{v} 1371\\
T83  & \ion{Mg}{v} 351, 352, 353, 354, 355; \ion{Ne}{v} 480, 481, 483;\\
     & \ion{Ne}{vi} 559, 563; \ion{Ne}{v} 570, 572;\\
     & \ion{O}{v} 630, 759, 760, 761, 762; \ion{O}{vi} 1032, 1038;\\
T99  & \ion{Mg}{v} 312; \ion{Mg}{vi} 350, 351; \\
     & \ion{Mg}{v} 351, 352, 353, 354, 355; \\
     & \ion{Mg}{vi} 389, 390, 399, 401, 403; \ion{Ne}{vi} 452, 453, 454; \\
     & \ion{Ne}{vii} 465; \ion{Ne}{vi} 559, 563; \ion{O}{vi} 1032, 1038 \\
T111 & \ion{Mg}{vi} 389, 390, 399, 401, 403; \ion{Ne}{vi} 559, 563; \\ 
     & \ion{Ne}{vii} 465, 559, 560, 561, 562, 563, 565\\
T116 & \ion{Mg}{vii} 364, 365, 368; \ion{Mg}{vi} 389, 390, 399, 401, 403; \\
     & \ion{Mg}{vii} 429, 431, 435; \ion{Ne}{vii} 465; \ion{Ne}{vi} 559, 563;\\
     & \ion{Ne}{vii} 559, 560, 561, 562, 563, 565; \ion{Ne}{viii} 770, 780\\
T117 & \ion{Fe}{ix} 171; \ion{Si}{vi} 246, 249; \ion{Mg}{vii} 364, 365, 368;\\
     & \ion{Mg}{vi} 389, 390, 399, 401, 403; \ion{Mg}{vii} 429, 431, 435; \\
     & \ion{Ne}{vii} 465; \ion{Ne}{vi} 559, 563;\\
     & \ion{Ne}{vii} 559, 560, 561, 562, 563, 565; \ion{Ne}{viii} 770, 780;\\
     & \ion{O}{vi} 1032, 1038\\
T112 & \ion{Fe}{ix} 171; \ion{Mg}{vii} 364, 365, 368; \\
     & \ion{Mg}{vi} 389, 390, 399, 401, 403; \ion{Ne}{vii} 465; \\
     & \ion{O}{vi} 1032, 1038\\
\hline
\end{tabular}
\end{table}

In Table~\ref{vitcar} we list the most prominent metallic wind lines in
predicted spectra that can be used to search for wind signatures in CSPN
spectra. As the wind becomes more ionized for stars with higher effective
temperatures, higher ions become visible in the spectra and the most prominent
wind features shift toward lower wavelengths.

\section{Comparison with observations}

Before Gaia Data Release 2 \citep[hereafter DR2]{gaia2}, the observational
estimates of CSPN mass-loss rates were problematic due to poorly known
distances. To overcome this, \citet{btpau} used hydrodynamical wind models to
fit the wind line profiles and to estimate the stellar radius (or luminosity)
and mass. On the other hand, \citet{herbian} provide two sets of stellar and
wind parameters; one was derived assuming the most likely stellar mass of CSPNe is about
$0.6\,M_\odot$ (adopted here) and second one was derived when using various estimates for
distances.

\begin{figure}[t]
\centering \resizebox{\hsize}{!}{\includegraphics{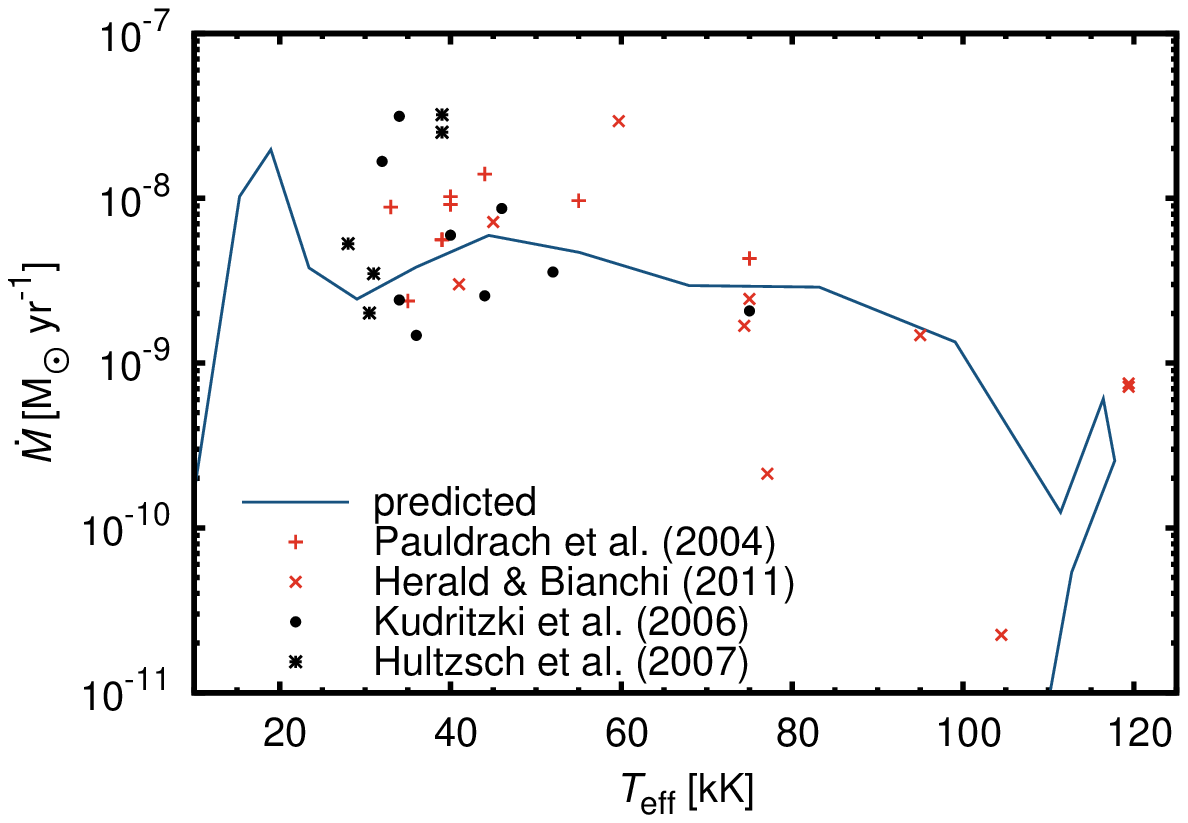}}
\caption{Predicted mass-loss rate as a function of stellar temperature (solid
blue line) in comparison with the observational results of \citet{herbian} based on
a UV analysis, the observational results of \citet{kuplan} and \citet{hulpul} based on
an optical analysis, and the predictions of \citet{btpau}.}
\label{dmdtt}
\end{figure}

In Fig.~\ref{dmdtt} we compare our predicted wind mass-loss rates with
observational and theoretical results for hydrogen-rich CSPNe as a function of
effective temperature. For a reliable determination of mass-loss rates from
observations, the inclusion of small-scale inhomogeneities (clumping) is
important \citep{hopak}. The estimates of \citet{btpau} neglect the influence of
clumping on the line profiles, while \citet{herbian}, \citet{kuplan}, and
\citet{hulpul} account for clumping. Because the luminosities of observed CSPNe
are generally higher than the luminosity assumed here and since the mass-loss rate
scales with luminosity on average as $\dot M\sim L^\alpha$, where $\alpha=1.63$
\citep{cmfkont}, we scaled the empirical mass-loss rates by a factor of
$(L_\text{CSPN}/L)^\alpha$, where $L_\text{CSPN}$ is the typical luminosity
assumed here (we adopted $\log(L_\text{CSPN}/L_\odot)=3.52$) and $L$ is the
stellar luminosity derived from observations. From Fig.~\ref{dmdtt}, it follows
that the mass-loss rates derived from literature slightly increase with
$T_\text{eff}$ up to roughly 50~kK, when they start to decrease. The predicted
mass-loss rates nicely follow this trend, albeit with much less scatter. Our
predicted mass-loss rates reasonably agree with the results of \citet{herbian}. Our
predictions are slightly lower than the rates derived by \citet{btpau}, who used
the Sobolev approximation to determine the line force. The Sobolev approximation
possibly leads to the overestimation of the mass-loss rates \citep[see discussion
in][]{cmf1}.

\begin{figure}[t]
\centering \resizebox{\hsize}{!}{\includegraphics{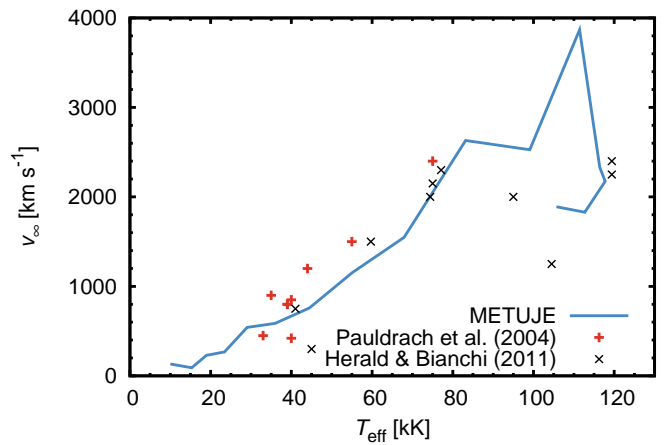}}
\caption{Predicted terminal velocity as a function of stellar temperature 
(solid blue line) in comparison with observational results.}
\label{vnekv}
\end{figure}

The wind terminal velocity increases during the post-AGB evolution with
increasing temperature, which is in agreement with observations (Fig.~\ref{vnekv}). These
variations stem from the linear dependence of the terminal velocity on the
escape speed \citep{cak,lsl,vysbeta}. If real CSPN radii are larger than assumed
here, as indicated by their larger luminosities derived from observations
(Table~\ref{pau}) and from evolutionary models \citep{milbe}, then the predicted
terminal velocities are in fact slightly lower than what was derived from
observations \citep[as shown by][]{btpau,btkas,hopak}. However, this problem is
not strong because the increase in the luminosity by 0.3\,dex would imply the
decrease in the wind terminal velocity by just 20\%. We face a similar problem
for our O star wind models \citep{cmfkont} where we attributed this problem
either to radial variations of clumping or to X-ray irradiation.

The observed terminal velocities show large scatter for $\Teff\gtrsim90\,$kK and
in some cases lie below the predicted values. This is possibly a result of lower
wind mass-loss rates and weaker line profiles, which do not trace the wind up to
the terminal velocity. In this case, the observational results can be considered
as lower limits. In addition, the stars with observed values of terminal
velocities do not need to be exactly on our selected evolutionary track.

\begin{figure}[t]
\centering
\resizebox{\hsize}{!}{\includegraphics{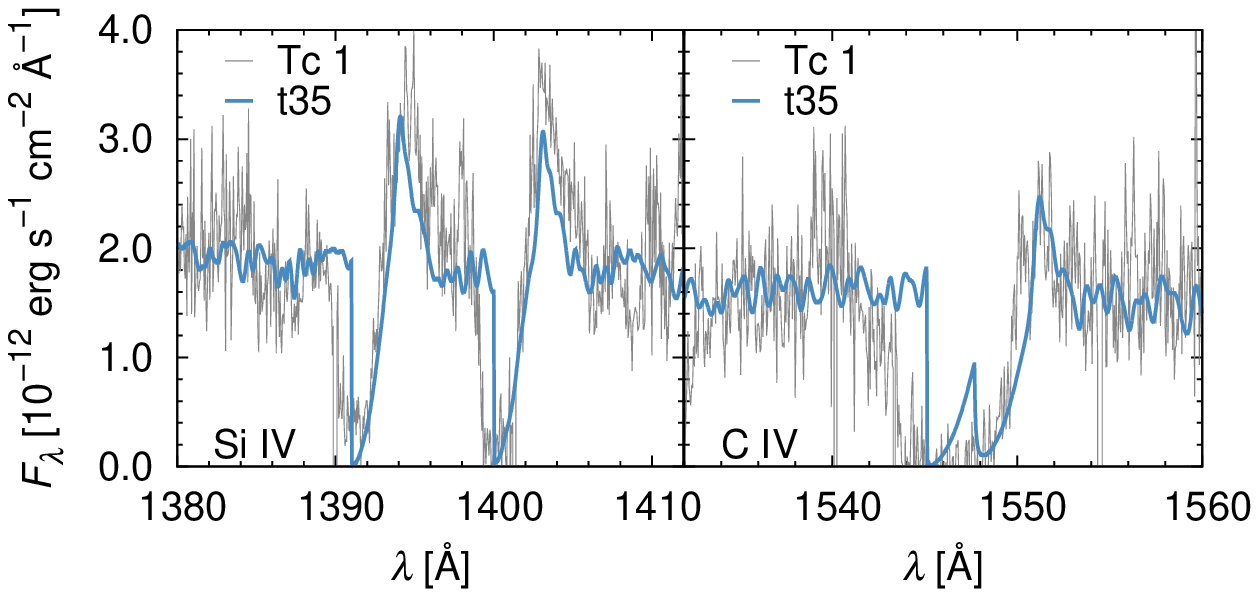}}
\resizebox{\hsize}{!}{\includegraphics{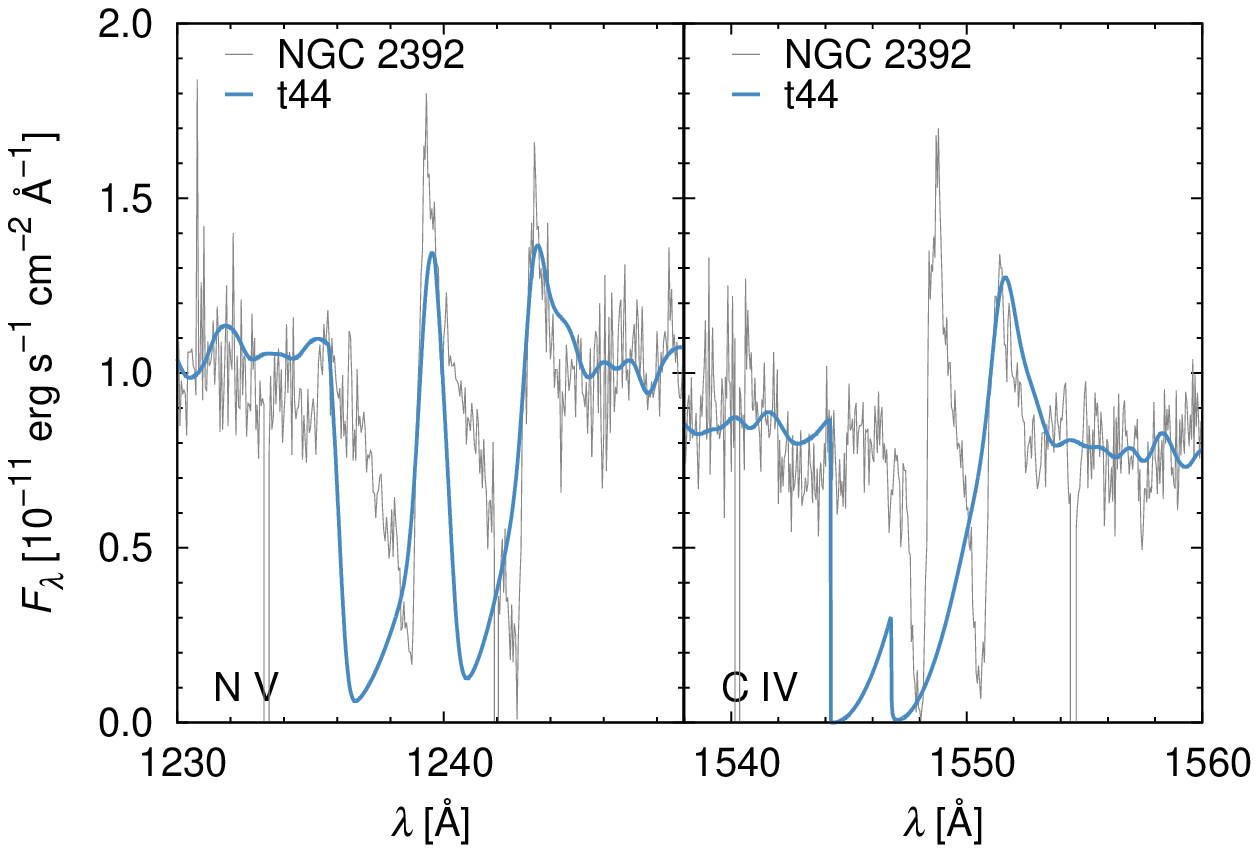}}
\resizebox{\hsize}{!}{\includegraphics{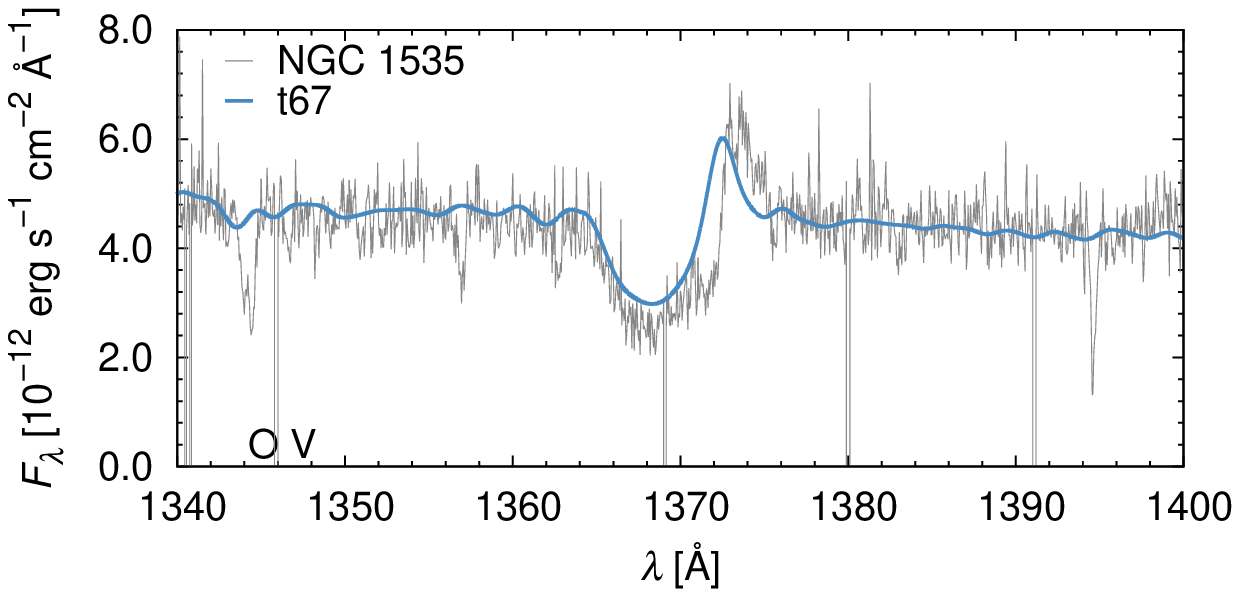}}
\caption{Comparison of predicted spectra from the grid in Table~\ref{hvezpar}
and IUE spectra of selected CSPNe. The predicted spectra were scaled to fit the
continuum. {\em Upper panel}: \object{Tc1} (central star
\object{HD 161044}) and IUE spectrum SWP 42675. {\em Middle panel}: \object{NGC
2392} and IUE spectrum SWP 19889. {\em Bottom panel:} \object{NGC 1535} and IUE
spectrum SWP 10821.}
\label{spektra}
\end{figure}

The ability of our code to predict reliable wind parameters is further
demonstrated in Fig.~\ref{spektra}, where we compare predicted spectra with
observed IUE spectra. We have not made any attempt to fit the spectra, we just
selected a model from the grid (Table~\ref{hvezpar}) with a temperature that is
closest to a given star. For the displayed stars, the strength of the emission
peak and the depth of the line profile are well reproduced by our models. On the
other hand, the extension of the blue part of the P~Cygni profiles is not fully
reproduced by the models. The profiles are in good agreement for the central star of
NGC~1535, while the extension of the absorption was overestimated for NGC~2392
and underestimated for Tc~1, implying problems with the terminal velocity estimation.

\begin{table*}[t]
\caption{Parameters of CSPNe with reliable distances.}
\label{pau}
\centering
\begin{tabular}{rcc@{\hspace{2mm}}c@{\hspace{2mm}}cc@{\hspace{2mm}}c@{\hspace{2mm}}cccccc}
\hline
Star & \multicolumn{4}{c}{\citet{btpau}}&&&\multicolumn{4}{c}{Gaia DR2}\\
& $T_\text{eff}$ & $\log g$ & $\log(L/L_\odot)$ & $M$ & $V$ & $E(B-V)$ &
$d$ & $\log(L/L_\odot)$ & $R$ & $M_\text{spec}$ & $M_\text{evol}$\\
& [K] & [cgs] && [$M_\odot$] & [mag] & [mag] & [pc] && [$R_\odot$]
& [$M_\odot$] & [$M_\odot$]\\
\hline
NGC 2392 & 40000 & 3.70 & 3.7& 0.41 &   9.68$^1$ & 0.03$^3$ & 2000 & $4.05\pm0.11$ & $2.2\pm0.4$ & $0.90\pm0.32$ & $0.66\pm0.06$\\
NGC 3242 & 75000 & 5.15 & 3.5& 0.53 &  12.15$^1$ & 0.04$^3$ & 1470 & $3.62\pm0.16$ & $0.4\pm0.1$ & $0.76\pm0.33$ & $0.55\pm0.03$\\
IC 4637  & 55000 & 4.57 & 3.7& 0.87 &  12.47$^2$ & 0.80$^2$ & 1330 & $3.90\pm0.09$ & $1.0\pm0.1$ & $1.31\pm0.41$ & $0.60\pm0.04$\\
IC 4593  & 40000 & 3.80 & 4.0& 1.11 &  10.84$^1$ & 0.07$^3$ & 2630 & $3.88\pm0.20$ & $1.8\pm0.4$ & $0.75\pm0.38$ & $0.59\pm0.09$\\
He 2-108 & 39000 & 3.70 & 4.2& 1.33 &  11.98$^1$ & 0.29$^2$ & 2790 & $3.71\pm0.13$ & $1.6\pm0.3$ & $0.45\pm0.17$ & $0.56\pm0.02$\\
IC 418   & 39000 & 3.70 & 4.2& 1.33 &  10.0$^2$  & 0.16$^3$ & 1550 & $3.83\pm0.10$ & $1.8\pm0.3$ & $0.60\pm0.20$ & $0.58\pm0.03$\\
NGC 6826 & 44000 & 3.90 & 4.2& 1.40 &   9.6$^1$  & 0.05$^3$ & 1580 & $4.02\pm0.10$ & $1.8\pm0.3$ & $0.90\pm0.29$ & $0.65\pm0.05$\\
\hline
\end{tabular}
\tablefoot{Source of photometric data: \tablefoottext{$^1$}{Simbad,}
\tablefoottext{$^2$}{\citet{menda}},\tablefoottext{$^3$}{\citet{prasan}}.}
\end{table*}

The wind terminal velocity scales with the escape speed, assuming smooth
spherically symmetric wind \citep{lsl}. This dependence enabled \citet{btpau} to
determine the stellar mass. However, the CSPN masses estimated using this method
fall significantly below or above the canonical mass of about $0.6\,M_\odot$
(see also Table~\ref{pau}). Our models face similar problems as shown in
Fig.~\ref{spektra} because our models underestimate the terminal velocity in
\object{Tc1} (as measured from the blue edge of P~Cygni profile) and
overestimate the terminal velocity in \object{NGC 2392}.

The wind terminal velocity is influenced by the ionization balance in the outer
wind and may be affected by such processes as clumping or shock X-ray ionization
\citep[e.g.,][]{irchuch}. The aim to reproduce the observed values of the
terminal velocity with our models would lead to higher surface gravities and
masses, similarly as in \citet{btpau}. To resolve this problem, we took
advantage of recent Gaia DR2 parallaxes \citep{gaia1,gaia2} to improve the
parameters of CSPNe \citep[see also][]{gonzasvatamari}. We determined the absolute magnitudes with visual magnitudes
$V$ from Simbad and \citet{menda}, extinction parameters taken either from maps
of \citet{prasan} or adapted from \citet{menda}, and distances from Gaia DR2
data. The emergent fluxes from our models
with $ 18\,\text{kK}< T_\text{eff}<100\,$kK, on average, give the bolometric
correction
\begin{equation}
\text{BC} = -3.73 -6.96 \zav{\log T_\text{eff}-4.64}-
 2.34 \zav{\log T_\text{eff}-4.64}^2,
\end{equation}
which was determined using Eq.~(1) of \citet{ostar2003} with $V$ response curves
derived from Mikul\'a\v sek (private communication). From this, we derived the
luminosity \citep[using a formula from][]{okali} with the effective temperature, the
stellar radius, and finally the mass using spectroscopic surface gravity.

The results given in Table~\ref{pau} do not show any large group of
near-Chandrasekhar mass CSPNe due to reduced luminosities compared to
\citet{btpau}. The average CSPN mass $0.81\pm0.18\,M_\odot$ reasonably agrees
with a typical mass of white dwarfs of about $0.59\,M_\odot$
\citep{malybilymuz}. Moreover, we determined the evolutionary mass of studied
stars from their luminosities and effective temperatures using post-AGB
evolutionary tracks of \citet[see also Table~\ref{pau}]{milbe}. The evolutionary
masses agree with spectroscopic ones within errors, albeit the spectroscopic
determination gives higher values on average. This may reflect a similar problem
termed a "mass discrepancy" found in O stars \citep{hekuku,marpula}. Nevertheless,
the average evolutionary mass of studied stars $0.60\pm0.03\,M_\odot$ perfectly
agrees with a typical mass of white dwarfs \citep{malybilymuz}; consequently, our
results point to a nice agreement for all mass determinations of post-AGB stars.

\begin{figure}[t]
\centering \resizebox{\hsize}{!}{\includegraphics{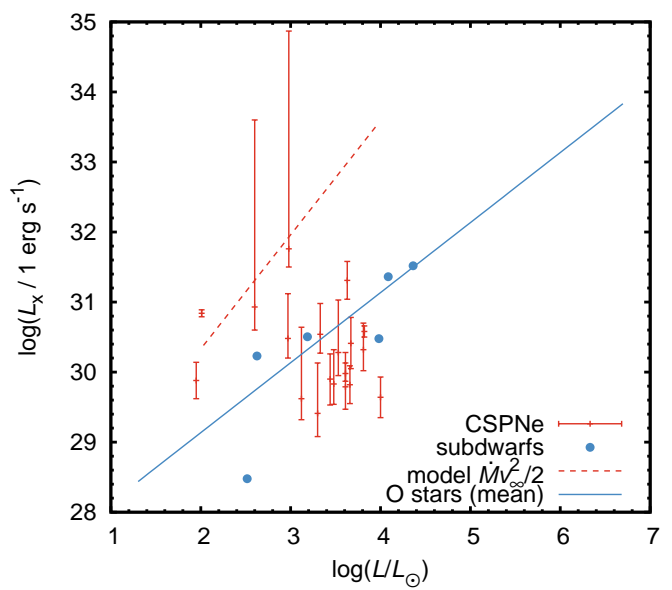}}
\caption{Relation between observed X-ray luminosities and bolometric
luminosities for CSPNe \citep{monty} and subdwarfs
\citep{dvoj9,dvoj5,bufacek,sam11,dvoj18}. We overplotted the mean relation for O
stars \citep[blue solid line]{naze} and the wind kinetic energy loss per unit of
time (red dashed line; derived using $v_\infty=1000\,\kms$ and mass-loss rate
from Eq.~\eqref{dmdtcspn} for $T=100\,$kK).}
\label{lxlbolpn}
\end{figure}

Some planetary nebulae are sources of X-ray emission, which may have a diffuse
component coming from the nebula \citep[e.g.,][]{chu,pekelnik} and a point-source
component associated with the central star \citep{svobodnik,monty}. X-ray
emission of O stars, which is supposed to originate in their winds
\citep{felpulpal}, shows a strong correlation between the X-ray luminosity and
stellar luminosity roughly as $\lx\approx10^{-7}L$ \citep{naze}. We compared the
relation between the X-ray luminosity and the stellar luminosity of CSPNe
derived from Chandra data \citep{monty} with corresponding relations obtained
for O stars and subdwarfs (Fig.~\ref{lxlbolpn}). All of the relations roughly follow
the same trend. Moreover, the observed X-ray luminosities are, in most
cases, lower than the wind kinetic energy lost per unit of time $\dot M
v_\infty^2/2$. This shows that the point-source X-ray luminosities of most of
CSPNe are consistent with their origin in the wind and that the wind is strong
enough to power the X-ray emission.  The only outlier in Fig.~\ref{lxlbolpn} is
\object{LoTr 5} for which the X-ray emission likely originates in the corona
around the late-type companion \citep{dvoj9}.

\section{Influence of metallicity}
\label{inmet}

\begin{table}
\caption{Predicted wind parameters for individual metallicities.}
\centering
\label{hvezparz}
\begin{tabular}{r*{3}{@{\hspace{1.5mm}}c@{\hspace{1.5mm}}r}}
\hline
\hline
Model&\multicolumn{2}{c}{$Z_\odot/3$}&\multicolumn{2}{c}{$Z_\odot$}&
\multicolumn{2}{c}{$3Z_\odot$}\\
& $\dot M$ & \vnek & $\dot M$ & \vnek & $\dot M$ & \vnek \\
\hline
T15  & $4.1\times10^{-9}$ &  40 & $1.0\times10^{-8}$ &  90 & $1.2\times10^{-8}$ &  100\\
T18  & $7.0\times10^{-9}$ & 180 & $2.0\times10^{-8}$ & 230 & $4.3\times10^{-8}$ &  90\\
T23  & $1.7\times10^{-9}$ & 240 & $3.8\times10^{-9}$ & 270 & $9.4\times10^{-9}$ & 300\\
T29  & $1.8\times10^{-9}$ & 490 & $2.4\times10^{-9}$ & 540 & $3.2\times10^{-9}$ & 520\\
T35  & $1.6\times10^{-9}$ & 550 & $3.8\times10^{-9}$ & 590 & $7.7\times10^{-9}$ & 550\\
T44  & $2.6\times10^{-9}$ & 880 & $6.0\times10^{-9}$ & 760 & $8.1\times10^{-9}$ & 780\\
T55  & $3.0\times10^{-9}$ & 980 & $4.7\times10^{-9}$ &1150 & $4.9\times10^{-9}$ &1150\\
T67  & $2.2\times10^{-9}$ &1340 & $3.0\times10^{-9}$ &1550 & $4.6\times10^{-9}$ &1420\\
T83  & $2.2\times10^{-9}$ &1280 & $2.9\times10^{-9}$ &2630 & $4.5\times10^{-9}$ &2640\\
T99  & $7.2\times10^{-10}$&1680 & $1.3\times10^{-9}$ &2530 & $2.3\times10^{-9}$ &2510\\
T116 & $1.2\times10^{-10}$&1200 & $6.1\times10^{-10}$&2330 & $7.4\times10^{-10}$&3020 \\
T117 & $2.9\times10^{-12}$&1520 & $2.6\times10^{-10}$&2170 & $3.0\times10^{-10}$&4520 \\
T112 & \multicolumn{2}{c}{no wind} & $5.4\times10^{-11}$&1830 & $1.4\times10^{-10}$ & 3370\\
T105 & \multicolumn{2}{c}{no wind} & $5.1\times10^{-13}$&1890 & $7.0\times10^{-11}$&2620  \\
\hline
\end{tabular}
\tablefoot{Mass-loss rates are given in units of \msr\ and terminal velocities
in units of \kms.}
\end{table}

Hot star winds are driven by a radiative force due to heavier elements;
consequently, the radiative force is sensitive to metallicity. Our calculations
assumed solar chemical composition, which might not be realistic in many cases.
The chemical composition of CSPNe partly reflects the composition of
the interstellar medium, which can be nonsolar depending on the environment. Massive
stars with an initial mass that is higher than about $1.2\,M_\odot$ experience third
dredge-up \citep{galino}, affecting the abundance of CNO elements and enriching
the abundance of s-process elements at the expense of iron \citep{ickos}.

We calculated additional wind models with an overabundance ($Z=3Z_\odot$) and
underabundance ($Z=Z_\odot/3$) of heavier elements. These might describe the
influence of the dredge-up phase and the initially metal-poor environment, respectively.
The resulting parameters of the models with nonsolar chemical composition are
given in Table~\ref{hvezparz}. On average, the wind mass-loss rate increases
with metallicity as $\dot M \sim (Z/Z_\odot)^{0.53}$, which is included in
Eq.~\ref{dmdtcspn}. The wind mass-loss rate is more sensitive to the metallicity for
low mass-loss rates $\dot M<10^{-9}\,\msr$ at which the wind is typically
driven by a few optically thick lines and, therefore, a small change in metallicity may cause a large change in the mass-loss rate.

\section{Influence of clumping}

\begin{table}
\caption{Predicted wind parameters for different clumping.}
\centering
\label{hvezparchuch}
\begin{tabular}{@{\hspace{0.5mm}}r*{3}{@{\hspace{1.5mm}}c@{\hspace{1.5mm}}r}@{\hspace{0.5mm}}}
\hline
\hline
Model&\multicolumn{2}{c}{$\cc=1$}&\multicolumn{2}{c}{$C_2=100\,\kms$}&
\multicolumn{2}{c}{$C_2=300\,\kms$}\\
& $\dot M$ & \vnek & $\dot M$ & \vnek & $\dot M$ & \vnek \\
\hline
T15&$1.0\times10^{-8}$&  90&$1.0\times10^{-8}$ & 100& $1.0\times10^{-8}$ & 100\\
T18&$2.0\times10^{-8}$& 230&$2.6\times10^{-8}$ & 190& $2.4\times10^{-8}$ & 240\\
T23&$3.8\times10^{-9}$& 270&$8.9\times10^{-9}$ & 480& $6.8\times10^{-9}$ & 430\\
T29&$2.4\times10^{-9}$& 540&$2.9\times10^{-9}$ & 570& $2.5\times10^{-9}$ & 570\\
T35&$3.8\times10^{-9}$& 590&$6.7\times10^{-9}$ & 500& $5.1\times10^{-9}$ & 600\\
T44&$6.0\times10^{-9}$& 760&$6.5\times10^{-9}$ & 810& $6.5\times10^{-9}$ & 780\\
T55&$4.7\times10^{-9}$&1150&$6.5\times10^{-9}$ & 710& $5.6\times10^{-9}$ & 870\\
T67&$3.0\times10^{-9}$&1550&$4.2\times10^{-9}$ &1400& $3.6\times10^{-9}$ &1560\\
T83&$2.9\times10^{-9}$&2630&$1.5\times10^{-9}$ &2990& $2.5\times10^{-9}$ &2170\\
T99&$1.3\times10^{-9}$&2530&$2.1\times10^{-9}$ &2610& $1.8\times10^{-9}$ &2650\\
T111&$1.2\times10^{-10}$&3870&$7.1\times10^{-10}$&2920&$9.8\times10^{-10}$&2380\\
T116&$6.1\times10^{-10}$&2330&$5.7\times10^{-10}$&2780&$5.8\times10^{-10}$&2790\\
T117&$2.6\times10^{-10}$&2170&$3.3\times10^{-10}$&2800&$3.3\times10^{-10}$&2800\\
\hline
\end{tabular}
\tablefoot{Mass-loss rates are given in units of \msr\ and terminal velocities
in units of \kms.}
\end{table}

We assumed spherically symmetric and stationary winds in our study. In reality,
stellar winds are far from this assumption. It is generally considered that
inhomogeneities on a small-scale (clumping) have the most significant effect on
the wind structure. Clumping in hot star winds influences the ionization
equilibrium \citep{hamko,bourak,martclump,pulchuch} and the radiative transfer
in the case of optically thick clumps either in continuum \citep{lidarikala} or
in lines \citep{lidaarchiv,chuchcar,sund,clres1,clres2}. The origin of clumping
is likely the line-driven wind instability \citep{lusol,ornest,felto}, which may
amplify the photospheric perturbations seeded by subsurface turbulent motions
\citep{felpulpal,cabra,jian} or which may be self-initiated in the wind
\citep{sundsim}.

In hot star winds, clumping affects not only the wind empirical characteristics,
but also wind parameters. Optically thin clumping leads to an increase in the
predicted mass-loss rates \citep{muij,irchuch}, which is in contrast to their influence
on the empirical mass-loss rates. On the other hand, optically thick clumps
counteract this effect. This has been found in a flow with a smooth velocity
profile \citep{muij} and the effect of optically thick clumps on the ionization
structure in the wind with a nonmonotonic velocity profile \citep{sundchuchmod}
would also likely lead to the same result.

We included clumping into our wind models \citep[for details see][]{irchuch}
assuming optically thin clumps. The clumping is parameterized by a clumping
factor $\cc={\langle\rho^2\rangle}/{\langle\rho\rangle^2}$, where the angle
brackets denote the average over volume. Always $\cc\geq1$, and $\cc=1$ corresponds
to a smooth flow. We adopted the empirical radial clumping stratification from
\citet{najradchuch}
\begin{equation}
\label{najc}
\cc(r)=C_1+(1-C_1) \, e^{-\frac{\vel(r)}{C_2}},
\end{equation}
which gives $\cc(r)$ in terms of the radial wind velocity $\vel(r)$. For our
calculations, we inserted the radial velocity from the models without clumping
into Eq.~\eqref{najc}. Here, $C_1$ is the clumping factor that is close to the star and
the velocity $C_2$ determines the location of the onset of clumping.
\citet{najradchuch} introduced additional constants $C_3$ and $C_4$ to account
for the decrease of clumping in outer regions of dense winds \citep{pulchuch}.
We neglected this effect because we are mostly concerned with the behavior of
the wind that is close to the star. As a result, the formula becomes similar to what was
used by \citet{bouhil}, for example. Motivated by typical values derived in
\citet{najradchuch}, we assume $C_1=10$ and two values of $C_2$, namely
$C_2=100\,\kms$ and $C_2=300\,\kms$, for which the empirical H$\alpha$ mass-loss
rates of O stars agree with observations \citep{cmfkont}. This gives a slightly
higher value of the clumping factor than what was found from CSPN spectroscopy,
which is about $\cc\approx4$ \citep{kuplan,hopak}.

Wind parameters that were calculated with inclusion of clumping are given in
Table~\ref{hvezparchuch} in comparison with the parameters derived for smooth
winds ($\cc=1$). In general, with increasing clumping, the rate of processes that
depend on the square of the density also increases. Consequently, wind becomes less
ionized and the mass-loss rate increases because ions in the lower degree of ionization
accelerate wind more efficiently \citep{muij}. The effect is weaker for a higher
value of $C_2$ because with the increasing onset of clumping, the radiative force
becomes less affected in the region close to the star where the mass-loss rate
is determined. In exceptional cases, the mass-loss rate may decrease due to
clumping. This happens when the lower ions are able to drive the wind less
efficiently. For example, in the case of the T116 model, the clumping leads to
a change in the dominant ionization stage of magnesium from \ion{Mg}{vii} to
\ion{Mg}{vi}, which accelerates the wind less efficiently, thereby leading to
a lower mass-loss rate (see Table~\ref{hvezparchuch}).

Typically, the increase in the mass-loss rate is weaker than the dependence
$\dot M\sim\cc^{0.2-0.4}$ derived for O stars \citep{muij,irchuch}. In a present
sample, the mass loss rate typically varies as $\dot M\sim\cc^{0.1-0.2}$. The
reason for this weaker dependence on clumping compared to O stars is most likely
a lower contribution of iron lines to the radiative driving. The CSPN winds are
mostly driven by elements that are lighter than iron. These elements effectively have a
lower number of lines than iron, but these lines are significantly optically
thick and they stay optically thick even after the change in ionization due to
clumping. Because the radiative force due to optically thick lines is related to
their number and not to the number density of corresponding levels, the
radiative force only varies weakly with clumping.

Some models deviate more significantly from the mean increase in the mass-loss
rate with clumping. The model T23 with $\cc=1$ (i.e., for a smooth wind without
clumping) shows an increased mass-loss rate due to the commencing effect of the
bistability jump. With clumping, the fraction of \ion{Fe}{iii} becomes even
larger and the increase in the mass-loss rate with respect to the unclumped model is
more significant than for other models. The influence of clumping is also stronger
for winds at low mass-loss rates, which are typically given by a few lines
and which are sensitive to a small change in the radiative force.

The model T18 is only very weakly sensitive to clumping. If this is also true
for B supergiants, then this effect can partly explain the discrepancy
\citep{vysbeta,kuraci,hau} between theoretical and empirical mass-loss rates for
these stars in the region of the bistability jump. The same level of clumping at
both sides of the jump would then lead to a larger increase in the mass-loss
rate at the hot side of the jump, improving the agreement between theory and
observation. The influence of clumping on the radiative transfer in the
optically thick H$\alpha$ line \citep{petchuch} or different growth rates of
instabilities \citep{drichuch} can also contribute to explain the discrepancy.

\section{Discussion}

\subsection{Influence of the magnetic field}

White dwarfs, which are the descendants of central stars of planetary nebulae, are found
to be magnetic \citep[see][for a review]{ab} with surface field strengths
ranging from a few kilogauss up to about 1000~MG \citep[e.g.,][]{valy,asb,landbt}. From
this, we can expect strong surface magnetic fields, even in CSPNe. However, the
observations show that this is not the case, placing the upper limit of the
magnetic field strength to about 100~G \citep{prejdi,ara,leo,stefi}. This might
pose problems for the models describing shapes of some planetary nebulae by
magnetic fields \citep{magplan1,magplan2,magplan3,magplan4}. On the other hand,
the detection of few G fields in post-AGB stars \citep{nasli} shows that the
magnetic fields with strength of the order 10~G may still be present in CSPN. In
such a case, it is likely that the fields are frozen in the stellar plasma
(similar to presumable fossil fields in hot main-sequence magnetic stars) and
that the fields are stable over the CSPN lifetime. We discuss the effect of such
fields on line driven winds.

The effect of the magnetic field is given by the ratio of the magnetic field
density and the wind kinetic energy density: ${\frac{1}{8\pi}B^2}/({\frac{1}{2}\rho
v^2})$. With a wind mass-loss rate of $\dot M=4\pi r^2\rho v$ and by replacing
$B=B_\text{eq}$ with the equatorial field strength, $r=R_\ast$ with the stellar
radius, and $v=v_\infty$ with the terminal velocity, we derive
\begin{equation}
\label{eta}
\eta_\ast=\frac{B_\text{eq}^2R_\ast^2}{\dot Mv_\infty},
\end{equation}
which is the wind magnetic confinement parameter \citep{udo}. For weak
confinement, $\eta_\ast\lesssim1$, the wind energy density dominates over the
magnetic field energy density and the wind speed is higher than the Alfv\'en
velocity. The magnetic field opens up and the wind flows radially. For strong
confinement, $\eta_\ast\gtrsim1$, the magnetic field energy density dominates
over the wind energy density. Wind is trapped by the magnetic field and
collision of wind flow from opposite footpoints of magnetic loops leads to
magnetically confined wind shocks \citep{bamo}.

In rewriting Eq.~\eqref{eta} in terms of typical parameters of CSPNe, we derive
\begin{equation}
\label{peta}
\eta_\ast=7.7\zav{\frac{B_\text{eq}}{100\,\text{G}}}^2
\zav{\frac{R}{1\,R_\odot}}^2
\zav{\frac{\dot M}{10^{-9}\,M_\odot\,\text{yr}^{-1}}}^{-1}
\zav{\frac{v_\infty}{10^3\,\kms}}^{-1}.
\end{equation}
Consequently, even the magnetic field with strength that evades detection on the
order of $10\,$G can still significantly affect the stellar wind especially in
early post-AGB evolutionary stages with a large radius or in later phases with a low
mass-loss rate.

The case of strong confinement $\eta_\ast\gtrsim1$ leads to wind quenching
\citep[e.g.,][]{magvln}. One can distinguish between two types of loops, closed
and open ones. Within a closed loop, the wind speed is always lower than the
Alfv\'en speed and wind confined within a closed loop never leaves the star
\citep{bamo}. The wind velocity at the apex of the last closed loop is just
equal to the Alfv\'en speed. The wind only leaves the star on open loops on
which the wind velocity exceeds the Alfv\'en speed at some point. Because the
open loops cover just a fraction of the stellar surface, this leads to a reduction
in the mass-loss rate, that is, to the wind quenching. Assuming a dipolar field
$r=R_\text{apex}\sin^2\theta$, where $R_\text{apex}$ is the apex radius and
$\theta$ is the colatitude, equating $R_\text{apex}=R_\text{A}$ gives the
magnetic field that is open for $\theta<\theta_\text{A}$, given by \citep{udo}
\begin{equation}
\label{uhel}
\theta_\text{A}=\arcsin\sqrt{\frac{R_\ast}{R_\text{A}}},
\end{equation}
where ${R_\text{A}}/{R_\ast}\approx0.3+(\eta_\ast+0.25)^{1/4}$ is the Alfv\'en
radius \citep{udorot}. Numerical simulations \citep{udorot} give a slightly lower
maximum radius of a closed magnetic loop as $R_\text{c}\approx
R_\ast+0.7(R_\text{A}-R_\ast)$. In the case of strong confinement, the wind
only leaves the star in the direction of magnetic poles for angles
$\theta<\theta_\text{A}$, forming a jet-like structure. The wind, therefore, only
interacts with material from previous evolutionary stages in two opposite
directions. This could possibly lead to a bipolar structure of some planetary
nebulae \citep[e.g.,][]{kohout,bipol1,bipol2}.

Mass-loss via magnetised stellar wind also leads to the angular momentum loss
and to the rotational braking \citep[e.g.,][]{sokarpnmag}. The spin-down time
depends on the stellar and wind parameters, on the magnetic field strength, and
on the moment of inertia constant $k$ \citep{brzdud}. To obtain a rough estimate
of the moment of inertia constant $k$, which is not available from the
\citet{vasiwo} models, we used the MESA evolutionary code \citep{mesa1,mesa2}.
We let a star with an initial mass of $1\,M_\odot$ evolve until the white dwarfs stage
with a final mass of $0.52\,M_\odot$. The moment of inertia constant
$k=3.3\times10^{-5}$ was derived from a model corresponding to a CSPN star with
$\Teff=15\,$kK. For typical stellar and wind parameters of CSPNe, Eq.~(25) in
\citet{brzdud} gives the spin-down time of about $10^3$\,yr for 100~G magnetic
field strength. This is an order of magnitude shorter than the typical CSPN
life-time \citep{milbe}. Consequently, magnetised CSPN significantly brake down
their rotation, providing an additional explanation for low rotational velocities of
white dwarfs \citep{hermelin}.

Magnetic braking at the top of the white dwarf cooling sequence can explain
extremely long rotational periods observed in some strongly magnetised white
dwarfs \citep{balbrot}. With $k=0.12$, which was derived for
the hot white dwarf stage of the MESA evolutionary track
discussed in a previous paragraph,
with a typical mass-loss rate of
$10^{-10}\,M_\odot\,\text{yr}^{-1}$, and with $B=100\,$MG, the spindown time is
about $10^4$\,yr \citep{brzdud}, which is a few orders of magnitude shorter than
the white dwarf cooling timescale.

%

Magnetic field only influences the mass-loss via its channeling along the
magnetic field lines, while the influence of the Zeeman effect is negligible for
magnetic fields that are weaker than about $10^5\,$G \citep{malabla}. On the other hand,
some white dwarfs show a magnetic field that is stronger than this limit. Consequently,
for hot white dwarfs with extremely strong magnetic fields, the line splitting
due to the Zeeman effect might affect the line force and the mass-loss rate.

\subsection{Shaping a planetary nebulae}

Much effort was put into understanding the shapes of planetary nebulae since
the proposition of \citet{kwok} that these nebulae originate from the collision
of fast weak wind with slow dense outflow coming from previous evolutionary
phases. It is typically assumed that the AGB wind is the main culprit of
the aspherical shape of many planetary nebulae, but the wind of CSPNe may also play
its role. There are several effects that lead to departures of the CSPN wind
structure from spherical symmetry and which may therefore cause asymmetry of
planetary nebulae.

Magnetic fields have long been suspected to cause the bipolar shape of planetary
nebulae \citep[e.g.,][]{magplan3,magplan4}. We find that the magnetic
fields with strengths on the order of 10\,G that evade the detection may still
have a significant impact on the wind (see~Eq.~\eqref{peta}). A magnetic field with
$\eta_*\gtrsim1$ (introduced in Eq.~\eqref{eta}) confines the wind into a form
of narrow outflow with an opening angle given by the Alfv\'en radius
Eq.~\eqref{uhel}. The magnetic field axis may likely be tilted with respect to
the rotational axis as in magnetic main-sequence stars \citep[see][for a recent
review]{wadne}. In such a case, the narrow outflow (jet) precesses with
the rotational period. Such precessing jets were inferred from observations
\citep{sahaj}. In the case of fast rotation, the rotational signature is likely
smeared out and the precessing jets may create just bipolar lobes.

Fast rotation is another mechanism that can potentially lead to the winds that
are not spherically symmetric. As a result of fast rotation, the stellar equator
becomes cooler, leading to weaker flow in the equatorial region than in the polar
region \citep{bezdisko}. Such effects are important when the rotational velocity
is close to the critical rotational speed $v_\text{crit}=\sqrt{2GM/(3R_*)}$,
which is $v_\text{crit}=60\,\kms$ for T10 model, but is
$v_\text{crit}=330\,\kms$ for T55 model, which roughly corresponds to typical
CSPN parameters. Due to a huge increase in the moment of inertia in late phases of
stellar evolution, angular momentum loss, and efficient coupling between the
core and envelope \citep{subtrot,gabtrot}, the stellar remnants are predicted to
rotate very slowly. Even magnetic torques are needed to reproduce the observed
rotational velocities of white dwarfs, which are typically rather slow
\citep{sklarny}. Consequently, even CSPNe with extreme rotational speeds rotate
at a small fraction of the critical velocity \citep{primac} and thus shaping of
planetary nebula by rotating post-AGB winds is unlikely.

Another effect that can influence the structure of planetary nebulae is the
bistability jump around $\Teff=20\,$kK (see Fig.~\ref{dmdtt}). During the
post-AGB evolution, the mass-loss rate decreases by a factor of about 8 and the
wind terminal velocity increases by a factor of roughly 2 around the bistability
jump. Therefore, after the star passes through the bistability jump, the fast
weak wind overtakes the slow dense wind, resembling the situation just after leaving
the AGB phase. This could possibly lead to the creation of an additional inner
shell of planetary nebulae.

\subsection{Wind limits}

The wind mass-loss rate is on the order of $10^{-9}\,M_\odot\,\text{yr}^{-1}$ during
the post-AGB evolutionary phase with constant luminosity. The mass-loss rate
starts to significantly decrease when the star turns right in the HR diagram and
settles on the white dwarf cooling track (Fig.~\ref{dmdtt}). For our selected
evolutionary track, there is a wind limit at about $105\,$kK below which there
are no homogeneous (i.e., hydrogen and helium dominated) stellar winds (see
Table~\ref{hvezpar}). The location of the wind limit nicely agrees with results
of \citet{unbu}, who found the wind limit at about $120\,$kK for solar abundance
stars with $\log g=7$ (see their Fig.~1).

The location of the wind limit was derived from additional models with a fixed
hydrodynamical structure, where we compared the radiative force with the gravity
for several different mass-loss rates between $10^{-16}\,\msr - 10^{-10}\,\msr$
\citep[for details see][]{metuje}. In all models descending on the white dwarf
cooling track with $\Teff<105\,$kK and $\log(L/L_\odot)<2.4,$ the radiative force
was lower than the gravity force. This means that the radiative force is not
able to drive a homogeneous wind and that there is no (hydrogen or helium
dominated) wind below $\Teff<105\,$kK.

The existence of the wind limit around $\Teff\approx105\,$kK is supported by
the observation of hot white dwarfs with near-solar chemical composition
\citep{wrk17,wrk18,wrr}. Despite their high gravity, the stellar wind possibly
prevents the gravitational settling \citep{unbu}. This region in the HR diagram
with fading wind where the gravity and radiative force compete can be
theoretically appealing. Similar processes as in classical chemically peculiar
stars, including rotationally modulated spectroscopic and photometric variability
\citep[e.g.,][]{kocuvir,prvalis}, may also appear in hot white dwarfs.

\citet{unbu98,unbu} showed that the wind mass-loss rate limit that prevents
abundance stratification is about $10^{-12}\,\msr$. From Table~\ref{hvezpar}, it
follows that the predicted division between homogeneous and stratified
atmospheres lies at about $110\,$kK, which is in reasonable agreement with observations.
An exact location of this boundary is uncertain due to a large sensitivity of
the wind mass-loss rate on the stellar parameters, such as the effective temperature and
metallicity (see, e.g., Table~\ref{hvezparz}).

To better understand the influence of metallicity on the location of the wind
limit, we performed additional wind tests with a factor of 10 higher abundance
of iron and silicon. This leads to a\ shift in the wind limit of 5\,kK to about
100\,kK in the models with overabundant iron and to a shift of about 15\,kK to
89\,kK in the models with overabundant silicon. The location of the wind limit
is also significantly different in He-rich stars and in PG~1159 stars
\citep{unbu}.

There is a possibility to launch an outflow even below the wind limit. Such
an outflow is not chemically homogeneous and consists of individual radiatively
accelerated metallic ions \citep{babela}.

There is a relatively strong line driven wind, even for the coolest star in our
sample, with $T_\text{eff}=10\,$kK. This indicates that even cooler star likely
have line-driven winds. This is consistent with the observation of P~Cygni line
profiles in the B-type post-AGB stars \citep{sarkar}.

\subsection{Effects of a multicomponent flow and frictional heating}

Hot star winds are mostly driven by the radiative force acting on heavier
elements, while the radiative force on hydrogen and helium is negligible. The
momentum acquired from radiation is transferred by Coulomb collisions to hydrogen
and helium, which constitute the bulk of the wind. In dense winds, the momentum
transfer is very efficient; consequently, dense winds can be treated as a
one-component flow assuming equal velocities of all ions \citep{cak76}. However,
in low-density winds, the Coulomb collisions are less efficient, which leads to
the frictional heating and to decoupling of individual components
\citep{treni,kkii,op,ufo,un}.

The effects of a multicomponent flow become important when the star appears at
the white dwarf cooling track, decreasing its luminosity and mass-loss rate
\citep{kielbt}. Their importance can be estimated from the value of the relative
velocity difference $x_{h\text{p}}$ \citep[Eq.~(18) in][]{nlteii} between a
given element $h$ and protons. When the velocity difference is low,
$x_{h\text{p}}\lesssim0.1,$ the multicomponent effects are unimportant, while
for $x_{h\text{p}}\gtrsim0.1,$ friction may heat the wind, and for
$x_{h\text{p}}\gtrsim1$ the element $h$ decouples.

We calculated the relative velocity difference $x_{h\text{p}}$ according to
\citet{nlteii} for all heavier elements. In a given wind model, the relative
velocity difference increases with the radius due to decreasing density and reaches
maximum in the outer parts of the wind. Our calculations show that for high
luminosities $\log(L/L_\odot)\gtrsim2.9,$ the wind density is so high that the
maximum nondimensional velocity difference is low, $x_{h\text{p}}<0.1$, and
therefore the flow can be considered as a one-component one. For lower
luminosities, $\log(L/L_\odot)\lesssim 2.5$ (in the model T105), the maximum
nondimensional velocity difference is higher, $x_{h\text{p}}>0.1$; consequently,
the wind may be frictionally heated or the components may decouple. The elements
that may decouple are those which significantly contribute to the radiative
driving, which are Ne, Mg, and Si for parameters corresponding to the T105
model. The decoupling affects the wind terminal velocity and may even reduce the
bulk mass-loss rate if it appears close to the critical point, where the wind
velocity is equal to the speed of the \citet{abbvln} radiative-acoustic waves
\citep[see also][]{fero}.

The frictionally heated winds were connected with the presence of
ultrahigh-excitation absorption lines in the spectra of hot white dwarfs
\citep{uhl,wrk18}. In such a case, the optical depth of a given line should be
close to one to cause a considerable absorption. In neglecting the finite cone
effects and the population of the upper level, the condition for unity Sobolev line
optical depth in a supersonic wind \citep{casrp} reads as
\begin{equation}
\label{tau}
\frac{\pi e^2}{m_\text{e}\nu_{ij}}n_if_{ij}
\zav{\frac{\de v}{\de r}}^{-1} \approx1,
\end{equation}
where $n_i$ is the number density of atoms in the lower level of a corresponding
line transition with frequency $\nu_{ij}$ and oscillator strength $f_{ij}$ and
where $\de v/\de r$ is the velocity gradient. Assuming $f_{ij}\approx1$ and
approximating $\de v/\de r\approx v_\infty/R_*$, the minimum mass-loss rate
needed to cause a considerable absorption is
\begin{multline}
\dot M\approx4\pi R_*^2 v_\infty n_\text{H}m_\text{H}=
\frac{4m_\text{e}m_\text{H}}{e^2}\frac{\nu_{ij}R_*v_\infty^2}{Z_\text{el}}=\\
3\times10^{-18}\msr\frac{1}{Z_\text{el}}
\zav{\frac{\nu_{ij}}{10^{15}\,\text{Hz}}}
\zav{\frac{R_*}{0.01\,R_\odot}}\zav{\frac{v_\infty}{10^3\,\kms}}^2,
\end{multline}
where $Z_\text{el}$ is the ratio of number densities of a given element and
hydrogen. This shows that for CNO elements with $Z_\text{el}\approx10^{-3}$, the
absorption line profiles can be observed down to the bulk mass-loss rates of the
order $10^{-14}\,\msr$ in hot white dwarfs. Moreover, the wind with velocity on
the order of $10^3\,\kms$ has a high enough kinetic energy that just part of it is
able to heat the wind to the temperatures required to generate
ultrahigh-excitation lines. Therefore, the presence of ultrahigh-excitation
absorption lines can be possibly explained by the frictionally heated wind.

\section{Conclusions}

We studied line-driven winds of white dwarf progenitors along their evolution
between the AGB stage and the top of the white dwarf cooling track. We calculated global
wind models for individual stellar parameters obtained from evolutionary models
with a CSPN mass of $0.569M_\odot$ and predicted wind structure, including wind
mass-loss rates and terminal velocities. We compared derived results with
observations.

Line driven winds are initiated very early after leaving the AGB phase and
appear, at the latest, for $T_\text{eff}\approx10\,$kK (but most likely even earlier).
Because the wind mass-loss rate mostly depends on the stellar luminosity, which
is nearly constant along the evolutionary track, the mass-loss rate does not
vary significantly during most of the CSPN evolution and is on the order of
$10^{-9}\,\msr$. This nicely agrees with observations. We fit our resulting
mass-loss rate as a function of stellar effective temperature and provided
corresponding scalings that account for different luminosities and
metallicities. Line-driven winds fade out at the white dwarf cooling track for
temperatures of about $105\,$kK for solar metallicity composition. Possibly only
metallic winds exist below this limit.

Two features interfere with an otherwise nearly constant mass-loss rate along the
post-AGB evolutionary track. A bistability jump at around
$T_\text{eff}\approx20\,$kK due to the change in dominant iron ion from
\ion{Fe}{iii} to \ion{Fe}{iv} leads to a decrease in the mass-loss rate by a
factor of about 8 and to an increase in the terminal velocity by a factor of roughly 2
(during the evolution). At about $\Teff=45\,$kK, additional broad and weak
maximum of mass-loss rates appears, which we interpreted as a result of an increase
in the far-UV radiative flux.

Because the post-AGB evolution proceeds at roughly constant luminosity, the
stellar radius decreases with increasing effective temperature. This and the
proportionality between the escape speed and the wind terminal velocity cause
the wind terminal velocity to increase during the post-AGB evolution. This
nicely agrees with observed terminal velocities. The predicted and observational
values also agree, although we argue that this could be partly a consequence of
adopted luminosities, which are slightly lower than those derived from
observations of typical CSPNe and from updated evolutionary models
\citep{milbe}. Therefore, as in the case of O stars \citep{cmfkont}, our models
slightly underpredict the wind terminal velocities, which could be caused by
radial variations of clumping or by X-ray irradiation. As the winds start to
fade out close to the white dwarf cooling track, the wind terminal velocities
also decrease.

We calculated the number of ionizing photons and compared the values with
results of plane-parallel models. As a result, it follows that the number of
\ion{H}{i} ionizing photons can be derived from plane-parallel photospheric
models for $\Teff\gtrsim40\,$kK and that the plane-parallel models predict
a reliable number of \ion{He}{i} ionizing photons for $\Teff\gtrsim50\,$kK.  For
cooler stars, the effects of spherical symmetry and envelope extension are so
strong that the number of ionizing photons are not reliably predicted from
plane-parallel models. On the other hand, wind absorption above  the \ion{He}{ii}
ionization jump is so strong that the plane-parallel models never predict
the correct number of \ion{He}{ii} ionizing photons for any star with wind and even
results of spherical models are sensitive to model assumptions.

We discuss how the wind properties can contribute to the shaping of planetary
nebula. The bistability jump can lead to the appearance of an additional shell of
planetary nebula. Magnetic fields with strengths that are close to their observational
upper limits are still powerful enough to channel the wind along the field lines
and to affect the shaping of planetary nebulae \citep{stefi} and the rotational
evolution of CSPNe.

Our models provide reliable wind parameters that can be used in post-AGB
evolutionary calculations or in the studies of planetary nebulae.

\begin{acknowledgements}
This research was supported by grant GA\,\v{C}R 18-05665S. Computational
resources were provided by the CESNET LM2015042 and the CERIT Scientific Cloud
LM2015085, provided under the programme "Projects of Large Research,
Development, and Innovations Infrastructures". The Astronomical Institute
Ond\v{r}ejov is supported by a project \mbox{RVO:67985815} of the Academy of
Sciences of the Czech Republic.
\end{acknowledgements}

\bibliographystyle{aa}

\end{document}